\documentstyle[epsf,aps,pra,amssymb,multicol]{revtex}

\newcommand{\fig}[1]{Fig.\,\ref{#1}}
\newcommand{\ceq}[1]{Eq.\,({#1})}

\def\ground{a}
\def\excited{b}
\def\raman{c}
\def\i{i}
\def\j{j}

  \def\rexcited{|\excited\rangle}
  \def\lexcited{\langle\excited|}
  \def\rground{|\ground\rangle}
  \def\lground{\langle\ground|}
  \def\rraman{|\raman\rangle}
  \def\lraman{\langle\raman|}
  \def\ri{|\i\rangle}
  \def\li{\langle\i|}
  \def\rj{|\j\rangle}
  \def\lj{\langle\j|}

  \def\omexcited{\omega_{\excited}}
  \def\omground{\omega_{\ground}}

  \def\omraman{\omega_{\raman}}
  \def\omi{\omega_{\i}}
  \def\omj{\omega_{\j}}
  \def\omegaer{\omega_{\excited\raman}}
  \def\omegagr{\omega_{\ground\raman}}
  \def\omegair{\omega_{\raman\i}}

  \newcommand{\k}[1]{\underline{k}_{#1}}

  \def\R{\underline{r}}

  \def\deltag{\delta_{\ground}}
  \def\deltae{\delta_{\excited}}

  \def\gge{g}
  \def\ggr{g_{\ground}}
  \def\gir{g_{\i}}
  
  \def\ger{g_{\excited}}

  \def\Deltagr{\Delta_{\ground}}
  \def\Deltaer{\Delta_{\excited}}
  
  \def\Deltair{\Delta_{\i}}

  \newcommand{\ETA}[1]{\eta_{#1}}

\def\sm{\rground \lexcited}

  \def\a{\hat{a}}

  \def\x{\hat{x}}
  \def\px{\hat{p}_x}
  \def\dx{\Delta x_0}
  \def\y{\hat{y}}
  \def\dy{\Delta y_0}

  \newcommand{\A}[1]{\hat{a}_{#1}}
  \newcommand{\Ad}[1]{\left.{\hat{a}_{#1}^\dagger}\right.}

  \newcommand{\Nu}[1]{\nu_{#1}}

  \def\Tl{{T_{L}}}
  
  \def\tp{t^\prime}

  \def\basisp{|\psi_+\rangle}
  \def\basism{|\psi_-\rangle}
  \def\basispm{|\psi_\pm\rangle}
  \def\basismp{|\psi_\mp\rangle}

  \def\lbasispm{\langle\psi_\pm|}

  \def\sbasisp{|\psi_+\rangle_0}
  \def\sbasism{|\psi_-\rangle_0}
  \def\sbasispm{|\psi_\pm\rangle_0}

  \def\xbasisp{|\psi_+\rangle_x}
  \def\xbasism{|\psi_-\rangle_x}
  \def\xbasispm{|\psi_\pm\rangle_x}
  \def\xbasismp{|\psi_\mp\rangle_x}

  \def\ybasisp{|\psi_+\rangle_y}
  \def\ybasism{|\psi_-\rangle_y}
  \def\ybasispm{|\psi_\pm\rangle_y}
  \def\ybasismp{|\psi_\mp\rangle_y}

  \newcommand{\rtwomode}[2]{|{#1},{#2}\rangle}
  
  \newcommand{\rrotmode}[2]{|{#1},{#2}\rangle_{\it rot}}

  \newcommand{\rvib}[1]{|#1\rangle}

  \def\rstate{|\psi\rangle}
  \def\lstate{\langle\psi|}
  \def\rstatet{|\psi(t)\rangle}
  
  \def\rstatetp{|\psi(\tp)\rangle}
  \def\lstatetp{\langle\psi(\tp)|}

  \newcommand{\rpsi}[1]{|\psi_{#1}(t)\rangle}
  \newcommand{\lpsi}[1]{\langle \psi_{#1}(t)|}
  \newcommand{\rpsiT}[1]{|\psi_{#1}(\tau)\rangle}
  
  \newcommand{\rPsi}[1]{|\psi ({#1})\rangle}

  \newcommand{\rPsiev}[1]{|\psi^{#1}\rangle}
  \newcommand{\lPsiev}[1]{\langle \psi^{#1}|}

  \def\rstatee{|\psi_{\it el}\rangle}
  \def\lstatee{\langle \psi_{\it el}|}

  \def\rstatev{|\psi_{\it vib}\rangle}
  \def\lstatev{\langle \psi_{\it vib}|}

  \def\ro{\,\hat{\rho}\,}

  \newcommand{\jump}[1]{\hat{C}_{#1}}
  \newcommand{\jumpd}[1]{\hat{C}^\dagger_{#1}}
  \newcommand{\jumpp}[1]{\hat{C}^{\prime}_{#1}}

  \def\H{\hat{H}}
  \def\HO{\hat{H}_0}
  \def\HI{\hat{H}_{I}}
  \def\Heff{\hat{H}_{\it eff}}

  \newcommand{\U}[2]{\hat{U}({#1}-{#2})}
  \newcommand{\Ud}[2]{\hat{U}^\dagger({#1}-{#2})}
  \newcommand{\UU}[1]{\hat{U}({#1})}
  \newcommand{\UUd}[1]{\hat{U}^\dagger({#1})}

  \def\Hilbert{{\cal H}}
  
  \def\subHilbert{{\cal H}_0}

  \newcommand{\SubHilbert}[1]{{\cal H}_{#1}}

  \def\quantop{\hat{A}}
  \def\quantopd{\hat{A}^\dagger}
  \def\revop{\hat{U}}

  \def\Ecav{E_{C}}
  \def\Ecavamp{E_{C}}
  \def\kcav{k_{C}}
  \def\omcav{\omega_{C}}
  \def\gcav{g_{C}}

  \def\Elaser{E_{L}}
  \def\Elaseramp{E_{L}}
  \def\klaser{k_{L}}
  \def\omlaser{\omega_{L}}
  \def\glaser{g_{L}}

  \def\C{\hat{A}}
  \def\Cd{{\hat{A}}^\dagger}

\begin{document}
\draft

\title{Motional Quantum Error Correction}

\author{J.~Steinbach${}^\diamond$ and
J.~Twamley${}^{\star,\diamond}$}

\address{ ${}^\diamond$ Optics Section, Blackett Laboratory, Imperial College,
London SW7 2BZ, United Kingdom\thanks{Email:
j.steinbach@ic.ac.uk} }
\address{ ${}^\star$ Department of Mathematical Physics, National University of Ireland,
Maynooth, Co.\ Kildare, Ireland \thanks{Email:
jtwamley@thphys.may.ie}}
\vspace{.5cm}
\date{November 5, 1998}

\maketitle

\begin{abstract}
We examine the dynamics of a qubit stored in the motional degrees of
freedom of an ultra-cold ion in an ion trap which is subject to the
decoherence effects of a finite-temperature bath.  We discover an
encoding of the qubit, in two of the motional modes of the ion, which
is stable against the occurrence of either none or one quantum jump.
For the case of a zero-temperature bath we describe how to transfer
only the information concerning the occurence of quantum jumps and
their types
to a measuring apparatus, without affecting the ion's motional state
significantly.  We then describe how to generate a unitary restoration
of the qubit given the jump information, through Raman processes
generated by a series of laser pulses.
\end{abstract}

\begin{multicols}{2}

\section{Introduction}

Is it possible to preserve quantum coherence in systems that are
irreversibly coupled to the environment?  Apart from being of
fundamental interest, this question is of particular significance in
the area of quantum information processing \cite{quantcomp}, where the
first two-bit quantum gate operations have been demonstrated in
quantum optical systems such as trapped ions \cite{ionqc}, and optical
cavities \cite{qedqc}. In this context the above question has been
answered with ``yes'' through the developement of quantum error
correction concepts, in which quantum coherence is preserved by
exploiting multi-particle entanglement \cite{errorcorrect}. The
realization of quantum gates using nuclear magnetic resonance and
spin-spin coupling within a molecule \cite{nmrqc}, has already allowed
a test of simple ideas of such correction concepts \cite{cory98}.

At present, perhaps the most promising implementation of a quantum
information processor which comprises a small number of qubits is
based on a linear ion trap as first proposed by Cirac and Zoller
\cite{ioncomp95}. In most current formulations of such an ion trap
computer \cite{wineland98,hughes97}, the qubits are stored in two
electronic ground or metastable states of individual ions while the
quantum ``bus'' connecting the various qubits is mediated by the ion's
collective motional state either through the center-of-mass (CM) or
higher excited modes. At the same time, recent experiments have shown
that the decoherence effects in systems of trapped ions are dominated
by decoherence of the motional state \cite{wineland98,meekhof96}. This
is also supported by theoretical investigations
\cite{schneider98,mio98,heating}.

The aim of this paper is to explore possibilities for motional error
correction and to present a scheme which may be used to preserve
quantum coherences in the motional state of trapped ions. We consider
a single trapped ion whose motional degrees of freedom undergo
finite-temperature dissipative dynamics as described by a thermal
master equation \cite{louisell73}. In this situation we describe how
quantum information, when stored in the motional (bosonic) degrees of
freedom of a trapped ion, can be encoded in a way which allows its
active stabilisation. For the zero-temperature case we
explicitely demonstrate how such a stabilisation can be implemented in
a trapped-ion system. This stabilisation can correct for the loss of a
single excitation in the motional state and needs only a single ion
while making use of two of the three degrees of freedom of the ion's
motion. Such stabilisation techniques could be used to stabilise the
``bus'' from the effects of decoherence or, more speculatively, the
stored qubits in a bosonic quantum computer
\cite{BosonicQC}.

We will describe decoherence using the quantum jump approach
\cite{quantum-jump} and base our stabilisation scheme on the concept
of ``quantum jump inversion'' \cite{inversion}. To begin with, we
briefly review some of the underlying ideas and give an overview of
the method that we will use to stabilise the quantum information. In
Section \ref{trapping} we outline the physics of the motion of the
trapped ion while in Section \ref{dissipation} we go into detail
regarding the short-time dissipative evolution and the derivation of
the specific qubit encoding which allows the active stabilisation of
the stored quantum information. In Section \ref{detection} we describe
our method of periodically detecting whether a quantum jump has occurred 
and if
so, determine the type of jump. This step will involve a
projective measurement of the ion's electronic states which does not
significantly disturb the motion, a process which we discuss in
Section \ref{projection}.  Finally, in Section \ref{repair}, we show
how the unitary inversion of the detected decoherence effects (quantum
jumps) can be effected through adiabatic processes.

\subsection{Decoherence and Quantum Jump Inversion}
\label{decoherence}

Our scheme is based on the concept of ``quantum jump
inversion'' which was first presented by Mabuchi and Zoller
\cite{inversion} and represents a special example of the general
formalism of reversible quantum operations as studied in detail by
Nielsen and Caves \cite{Nielsen}. We will make some use of their
results in the following and briefly review them here, together with
the underlying description of decoherence through the formalism of
quantum trajectories \cite{carmichael93,monte-carlo} and the quantum
jump approach \cite{quantum-jump}.

In quantum optics, decoherence is described by coupling the system to
an infinitely large reservoir which models the environment. The
couplings between the system operators and the reservoir operators are
usually assumed to be weak and linear. This allows one to derive a
Markovian master equation for the dynamics of the system reduced
density operator $\ro,$ after tracing over the reservoir degrees of
freedom \cite{louisell73}. The general form of this master equation is
\begin{eqnarray}
 &&\frac{d}{dt} \ro = -\frac{i}{\hbar} [\H,\ro]
	\nonumber \\
 && - \sum_{m=0}^{N}
 \frac{1}{2} \left\{\jumpd{m}\jump{m}\ro + \ro\jumpd{m}\jump{m} -
 2\jump{m}\ro\jumpd{m} \right\}\,, \label{2}
\end{eqnarray}
where $\H$ is the system Hamiltonian and the operators $\jump{m}$
describe the different decay channels to the environment. The $\jump{m}$
originate from the linear coupling between system and environment and
are proportional to the system operators which couple to the
reservoir. 

An alternative approach to open quantum systems has been developed
which describes the system through an ensemble of pure state
evolutions through the formalism of quantum trajectories
\cite{carmichael93,monte-carlo}. In this method the system is
continuously monitored by performing measurements on the state of the
reservoir. Different measurement strategies lead to different
conditioned evolutions of the system which are termed
unravelings. More specifically, in the quantum jump method
\cite{quantum-jump}, one obtains an unraveling of the master equation
as arising from the conditioned evolution of the system where the
conditioning is supplied by detecting the loss of system excitations
through the counting of quanta in the environment. In the absence of any
measurement counts the system evolution is governed by the
non-Hermitian Hamiltonian
\begin{equation}
 \Heff = \H - \frac{i \hbar}{2} \sum_{m=0}^{N} \jumpd{m}\jump{m}\,,
 \label{4}
\end{equation}
which differs from the system Hamiltonian $\H$ due to the
information that we acquire from not observing any counts. On the other
hand, the information obtained from a detector count at a time $t$,
associated with the decay channel $m$, conditions the state of the
system according to
\begin{equation}
 \rPsi{t+} = \jump{m} \rPsi{t-},
 \label{6}
\end{equation}
where $\rPsi{t-}$ and $\rPsi{t+}$ describe the state of the system
immediately before and after the quantum jump, respectively.

In their article, Mabuchi and Zoller \cite{inversion} have formulated
conditions for the reversibility of quantum jumps through unitary
operations. We adopt here the more general notation of Nielsen and
Caves \cite{Nielsen} for the reversibility of quantum
operations. Nielsen and Caves have shown that, given the initial state
$\rstate$ of a quantum system is {\em known} to lie in a specific
subspace $\SubHilbert{0} \subset \Hilbert$, of the entire system
Hilbert space $\Hilbert,$ and the subspace $\SubHilbert{0}$ has the
property
\begin{equation}
 \lstate \quantopd \quantop \rstate = \mu^2\,, 
	\qquad \forall \rstate \in \subHilbert\,,
 \label{nccond}
\end{equation}
then the action of the quantum operation as described by $\quantop$ is
unitarily reversible. In other words, given \ceq{\ref{nccond}} then
there exists a unitary operation $\revop,$ such that $\revop \quantop
\rstate = \mu \rstate\,,\forall \rstate \in \subHilbert,$
where $\mu$ is a real constant \cite{Nielsen}. To illustrate the
condition in \ceq{\ref{nccond}}, consider a system with a single
bosonic degree of freedom, such as the electric field inside a
single-mode resonator, and where $\SubHilbert{0}$ is a two dimensional
subspace of $\Hilbert$ and is spanned by two basis states $\basisp$
and $\basism.$ If the quantum operation $\quantop = \a$, and describes
a quantum jump corresponding to the loss of a photon from the
resonator, then \ceq{\ref{nccond}} implies that {\em all\/} states
$\rstate \in \SubHilbert{0},$ possess the same mean cavity
excitation number. Alternatively, \ceq{\ref{nccond}} is satisfied if
the basis states $\basisp$ and $\basism$ are orthogonal, 
where $|\psi_\pm\rangle$ possess the same mean
excitation number {\em and\/} remain orthogonal after the quantum
jump. It is clear that in this situation the coherence present in the
subspace $\SubHilbert{0}$ (in the form of a coherent superposition of
the states $\basisp$ and $\basism$) is not damaged through the action
of the quantum operation $\quantop$ and it can be restored to the
state prior to the quantum jump through a unitary process. Mabuchi and
Zoller have exploited this fact to preserve quantum coherences in a
specially constructed situation in cavity quantum electrodynamics
(CQED) \cite{inversion}.
Following their work, Mensky \cite{mensky96} has studied the inversion
of quantum jumps under the aspect of measurement reversibility. More
recently, Vitali {\it et al.} \cite{vitali98} have proposed CQED
schemes which are similar in spirit to that of Ref.\cite{inversion} to
protect quantum states in cavities employing continuous feedback
methods.

\subsection{Stabilization Strategy}
\label{strategy}

In the following we outline how the strategy of quantum jump inversion
can be modified and implemented to stabilise a qubit which is stored
in the motional degrees of freedom of a trapped ion. The
implementation of a quantum jump inversion scheme requires {\em a
priori\/} knowledge about the decay channels that describe the effects
of decoherence as in \ceq{\ref{2}}. We will assume that the motional
degrees of freedom undergo finite-temperature dissipative dynamics
which are described by a thermal master equation
\cite{louisell73}. As we will show below, to derive a qubit
encoding which allows the unitary inversion of the quantum jumps
associated with this type of decoherence, it is sufficient to examine
the case of a zero-temperature bath. This will considerably simplify
our analysis. Irrespective of the particular choice of bath, the exact
inversion of a quantum jump will require: (i) A special choice of
motional subspace $\SubHilbert{0}$ in which to encode a qubit such
that the quantum jumps associated with a given type of decoherence are
unitarily reversible, (ii) a mechanism for detecting quantum jumps
without much disturbance of the motion, and (iii) a procedure which
accomplishes the unitary inversion of the detected decoherence
effect. While (i) is difficult, to satisfy (ii) and (iii) represents a
serious challenge for an operational quantum jump inversion strategy.

\noindent
\begin{minipage}{0.48\textwidth}
\begin{figure}[t]
    {
    \leavevmode
    \epsfxsize=80mm
    \epsffile{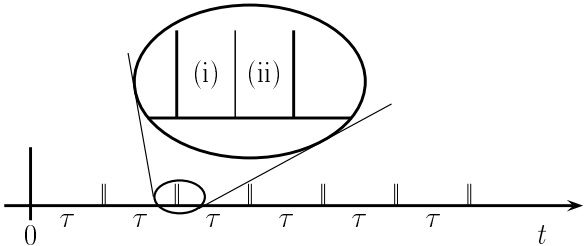}
    }
\caption{
Periodically performed projective measurements on the system: With a
period $\tau$ we interrogate the dissipative system evolution. Step
(i) serves to determine whether a quantum jump has occurred in the
preceding time interval $\tau.$ If a quantum jump is detected in step
(i) this is reversed through a unitary operation in step (ii) and the
quantum coherence in the system is restored to the original state.
}
\label{fig1}
\end{figure}
\vspace{1pt}
\end{minipage}

Before we go into more detail about how the above mentioned
requirements can be met in the physical situation of a trapped ion we
mark the following general considerations. The effects of motional
decoherence that have been observed in recent trapped ion experiments
are believed to be due to technical noise \cite{meekhof96}. In this
situation, even though one can obtain master equations for the
motional dynamics of the ions \cite{schneider98,mio98}, it is
difficult to isolate the particular environment to which the system is
coupled. Even in those cases where a physical quantum bath can be
identified, it is extremely difficult to monitor the system through
the quantum information transferred to the reservoir. The energies of
the motional excitations are too small to be easily detected. Thus, to
determine if a quantum jump has occurred we propose to {\em
periodically perform projective measurements} on the system directly
without gaining any information on the stored quantum superposition,
as illustrated in \fig{fig1}.  This is a significant difference to the
original formulation of a quantum jump inversion scheme
\cite{inversion}, where the dissipative system dynamics are {\em
continuously\/} monitored by imposing detectors directly on the
environment output channels and it makes our scheme more akin to
quantum error correction schemes \cite{errorcorrect}. The main
consequence of this modification lies in the fact that our quantum
jump detection scheme cannot reveal when in the time interval between
successive projective measurements the quantum jump occurred. We will
however still be able to unitarily invert the decoherence effects
caused by such a jump.

To deduce the proper encoding for the qubit we examine, in Section
\ref{dissipation}, the effects of motional decoherence associated with
a zero-temperature master equation given that the initial motional
state of the system $\rstatev$ lies in the subspace $\subHilbert
\subset \Hilbert$, which is spanned by two basis states $\basispm,$
and where $\Hilbert$ is the entire motional Hilbert space. As
mentioned above this is sufficient to obtain an encoding which
allows the unitary inversion of the quantum jumps associated with the
more general, finite-temperature reservoir. Based on the condition given in
\ceq{\ref{nccond}} we show that for the decoherence effects to be
unitarily reversible both in the case of a detected quantum jump, and
in the absence of a detection event, requires that the basis states
$|\psi_\pm\rangle$ comprise of at least two bosonic modes. We
therefore propose to encode the qubit as
\begin{equation}
 \rstatev = c_+ \basisp + c_- \basism\,,
 \label{encoding}
\end{equation}
in two of the ion's motional degrees of freedom. In this situation,
the effects of decoherence cause the system to evolve into a mixed
state
\begin{equation}
 \ro \approx p_0 \rPsiev{0}\lPsiev{0} + p_1 \rPsiev{1}\lPsiev{1} 
	+ p_2 \rPsiev{2}\lPsiev{2}\,,
 \label{8}
\end{equation}
on a time scale which is short compared with the characteristic decay
time. Here, $\rPsiev{0} \in \subHilbert$, $\rPsiev{1} \in
\SubHilbert{1},$ and $\rPsiev{2} \in \SubHilbert{2}.$ Since the projective
measurements which accomplish the detection of quantum jumps need to
discriminate between the three parts in the above mixture, we require
the three subspaces $\subHilbert,$ $\SubHilbert{1}$ and $\SubHilbert{2}$
to be mutually orthogonal. These constraints will lead us to the
choice
\begin{displaymath}
 |\psi_\pm\rangle=(|4,0\rangle+|0,4\rangle\pm \sqrt{2}|2,2\rangle)/2\,,
\end{displaymath}
for the basis motional states where $|n_x,n_y\rangle$ denotes a
two-dimensional motional Fock state. 

The detection of quantum jumps in the motional state of the trapped
ion represents step (i) in the schematic depiction of our
stabilisation scheme given in \fig{fig1}. This step consists of two
parts. First, a Raman induced process serves to entangle the mixed
state given above in \ceq{\ref{8}}, with orthogonal electronic states
of the ion, thereby transferring information about the occurrence of
quantum jumps into the electronic degrees of freedom.  Following this
entanglement operation, which we discuss in Section
\ref{entanglement}, we perform a measurement of the ion's electronic
degrees of freedom which projects the state from the mixture in
\ceq{\ref{8}} into a pure state. The projection could, in principle,
be done using quantum jump techniques
\cite{fluorescence,beige96}. However, in the situation where we aim to
protect the motional state of an ion from the effects of decoherence
we cannot tolerate the recoil from scattering a large number of
photons as this would lead to motional excitation and irreversibly
change the motional state \cite{decohereqc}. Instead, for the second
step of the quantum jump detection, we propose in Section
\ref{projection} an alternative projective measurement which is based
on the ``photon gun'' scheme discussed by Gheri {\em et al.\/}
\cite{Gheri}, where the ion is surrounded by a low-Q optical
cavity. Dependent on the ion's electronic state a laser pulse triggers
the transmission of a single-photon state out of the cavity where the
projection can be accomplished through photodetection. In the
Lamb-Dicke limit \cite{Lamb-Dicke,blockley92} we can practically
eliminate any motional recoil suffered by the ion while still
obtaining a signal whether a quantum jump has occurred or not. In
effect, we are coupling in another output channel to interrogate the
system in addition to those channels coupling the system to the
reservoir.

The unitary inversion of the detected quantum jumps which is
schematically depicted as step (ii) in \fig{fig1} represents perhaps
the most essential part of the stabilization scheme. This is addressed
in Section \ref{repair}. In the case where no quantum jump has been
detected the ion is left in its initial motional state as given in
\ceq{\ref{encoding}} and no further manipulation is required to
restore the quantum information to the original basis states. In the
case where a quantum jump has been detected, we describe, using two
adiabatic transfer processes and an intermediate resonant step, how
one can unitarily invert the decoherence effects associated with this
quantum jump. The involved processes have the advantage that they are
mostly achieved via adiabatic passage and thus are not subject to
noise due to spontaneous emission nor do they suffer from the strict
timing constraints imposed by resonant processes. The laser
arrangements needed to effect this unitary inversion of the detected
decoherence effects are not overly complex and may be experimentally
feasible.

In our analysis of the processes that constitute step (i) and (ii) in
the schematic depiction of our stabilisation scheme in \fig{fig1} we
will neglect all effects of decoherence during the corresponding
operations. In this situation, if the detection and inversion
processes are performed with a period $\tau$ which is rapid when
compared with the characteristic decoherence time $\gamma^{-1}$ the
probability for more than one quantum jump to occur in between two
successive projective measurements becomes negligible since this is
proportional to $(\gamma \tau)^2,$ \cite{carmichael93}. The detection
and unitary inversion of single quantum jumps then stabilises the
quantum information stored in the system.

The stabilisation scheme presented here allows the deterministic
(unitary) restoration of quantum information which is to be contrasted
with previous proposals where the correction contains probabilistic
(non-unitary) elements \cite{cirac96}.

\section{The trapping model}
\label{trapping}

Here we briefly introduce the trapping model that we use for the
analysis of our stabilisation scheme. This includes the description of
the quantised motion of the trapped ion, the relevant internal
electronic structure and the associated decoherence effects. 

\noindent
\begin{minipage}{0.48\textwidth}
\begin{figure}[h]
   {
    \leavevmode
    \epsfxsize=80mm
    \epsffile{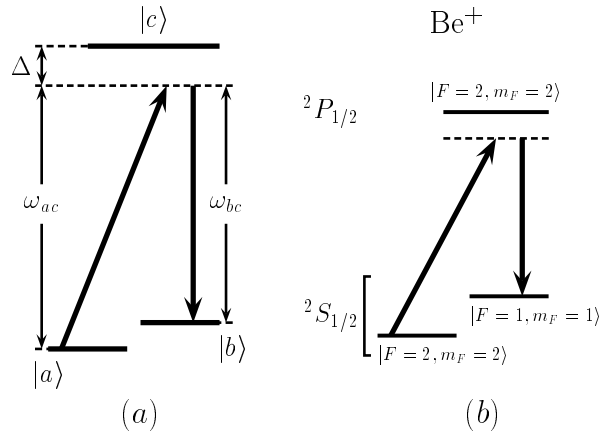}
   }
\caption{
(a) Three-level electronic configuration of the trapped ion considered
here. The two low lying states $\rground$ and $\rexcited$ are coupled
through a stimulated Raman interaction driven by two (correlated)
lasers at frequencies $\omegagr,\omegaer.$ The upper state $\rraman$
is detuned by a large enough amount $\Delta$ that it is never
appreciably excited. 
(b) Energy levels of a $\rm Be^+$ ion which realizes the effective
three-level configuration depicted in (a). The states $\rground$ and
$\rexcited$ are associated with the $\rvib{F=2,m_F=2}$ and
$\rvib{F=1,m_F=1}$ states within the fine-structure multiplet of the
$2s{}^2S_{1/2}$ ground state of $\rm Be$. The excited state
$\rraman$ is associated with the Zeeman-sublevel $\rvib{F=2,m_F=2}$
of the $2p{}^2P_{1/2}$ excited state [8].
}
\label{fig2}
\end{figure}
\vspace{1pt}
\end{minipage}

Consider the quantized CM motion of a single ion which is confined
within a trap potential that can be closely approximated as being
harmonic and which is characterized by the three frequencies of
oscillation, $\Nu{x}$, $\Nu{y}$, and $\Nu{z}$ along the $x$, $y$, and
$z$ directions \cite{bardroff96,harmonic}. As mentioned above we
propose to encode the qubit in the two bosonic degrees of freedom
which are represented by the ion's harmonic motion along the $x$ and
$y$ directions. We restrict our analysis to those two motional degrees
of freedom since the ion's motion along the $z$ direction plays no
role in our scheme. The only decoherence effect that we include in our
model is motional decoherence. To avoid any loss of quantum coherence
within the electronic degrees of freedom we require a scheme in which
spontaneous emission is highly suppressed. We therefore assume that
the relevant electronic degrees of freedom form a $\Lambda$ system as
depicted in \fig{fig2}(a), and that the electronic dynamics are confined
to the two states $\rground$ and $\rexcited$ which we choose to be two
ground state hyperfine levels. These are coupled by M1 and E2
transitions at best so that we can safely neglect any spontaneous
emission between those states \cite{bollinger91}.  At the same time
the $\rground \Leftrightarrow \rexcited$ transition can be driven by
two laser beams connecting the ground states $\rground$ and
$\rexcited$ through a common excited state $\rraman$ in a stimulated
Raman scheme \cite{monroe95,steinbach97}, which we briefly review in
Appendix \ref{ramanapp}. To give a specific example for the
realization of an effective three-level scheme as indicated in
\fig{fig2}(a) we have drawn in \fig{fig2}(b) the relevant energy level
structure of $\rm Be$ which, for example, was used in a series of
experiments in the group of Wineland at NIST
\cite{ionqc,wineland98,meekhof96}.

Because the trap potential is harmonic, the position and momentum
operators can be written in terms of creation and annihilation
operators for the trap quanta 
\begin{eqnarray*}
 \x &=& \sqrt{\frac{\hbar}{2\Nu{x}m}}\left(\A{x} + \Ad{x}\right)\,,
 \\
 \px &=& i \sqrt{\frac{\hbar\Nu{x}m}{2}}\left(\Ad{x}-\A{x}\right)\,,
\end{eqnarray*}
and similarly for $\y,$ and where $m$ is the mass of the ion. The free
Hamiltonian of the system then takes the form
\begin{equation}
 \HO = \sum_{\j=\ground,\excited,\raman} \hbar \omj \rj \lj 
 + \sum_{\j=x,y} \hbar \Nu{\j} 
	\left(\frac{1}{2} + \Ad{\j} \A{\j}\right)\,, 
 \label{20}
\end{equation}
where $\hbar \omground$ and $\hbar \omexcited$ denote the energy of
the two ground state hyperfine levels and $\hbar \omraman$ is the energy
of the electronic excited state. 

We assume that the motional degrees of freedom in the $x$ and $y$
directions undergo finite-temperature dissipative dynamics which are
described by a master equation of the form given in \ceq{\ref{2}}, and
where the associated decay channels are described by the operators
\begin{equation}
 \jump{x,y} = \sqrt{\gamma (\overline{n} + 1)}\,\A{x,y}\,, \quad
 \jumpp{x,y} = \sqrt{ \gamma \overline{n}}\,\Ad{x,y}\,,
 \label{60}
\end{equation}
and the quantity $\overline{n}$ gives the mean excitation number of
the reservoir \cite{louisell73}. Note that for simplicity we have
assumed the decay rate $\gamma$ to be the same for the $x$ and $y$
directions. In Section \ref{dissipation}, we show how a qubit can be
encoded in the motional degrees of freedom, such that the quantum
jumps associated with the decay channel operators given in
\ceq{\ref{60}} can be unitarily inverted. In Sections
\ref{detection} and \ref{repair} where we present our specific
proposal for the implementation of a stabilisation scheme in a
trapped-ion system we specialize to the zero-temperature case where
$\overline{n} = 0,$ and we are left with the channel operators
$\jump{x,y} = \sqrt{\gamma} \A{x,y}.$  We have chosen this simple
form of decoherence to exemplify our general strategy and to give an
explicit implementation for all the steps in our stabilisation scheme.

\section{Dissipative System Evolution and Stable Basis States}
\label{dissipation}

In this section we discuss the dissipative system evolution and, given
the form of the master equation as in \ceq{\ref{2}} together with the
channel operators in \ceq{\ref{60}}, deduce the orthogonal basis states $\basispm$
which allow the unitary inversion of the associated decoherence
effects. We refer to such states as {\em stable}. 

Before we go into more detail about the dissipative system evolution,
we note that for the derivation of stable basis states for a
finite-temperature master equation it is sufficient to consider the
zero-temperature case. The reason for this is that the unitary
reversibility of a quantum jump associated with the decay channel
$\jump{x}$ implies the unitary reversibility of a quantum jump through
the decay channel $\jumpp{x}$. This becomes clear from the following
argument. If a subspace $\SubHilbert{0}$ satisfies the condition given
in \ceq{\ref{nccond}} for the quantum operation $\quantop = \jump{x}$
which describes a quantum jump through the decay channel $\jump{x}$,
then the subspace $\SubHilbert{0}$ also satisfies the condition in
\ceq{\ref{nccond}} for the quantum operation $\quantop = \jumpp{x}$
associated with the decay channel $\jumpp{x}.$ This follows directly
from the form of the channel operators as in
\ceq{\ref{60}} and the commutation relation $[\A{x},\Ad{x}] = 1,$ for
the bosonic creation and annihilation operators. The same argument
holds for the decay channels $\jump{y}$ and $\jumpp{y}.$ In the
following we will therefore only consider the zero-temperature case
where the dissipative system dynamics are described by the master
equation 
\begin{equation}
 \frac{d}{dt} \ro = - \sum_{j=x,y}
 \frac{1}{2} \left\{\jumpd{j}\jump{j}\ro + \ro\jumpd{j}\jump{j} -
 2\jump{j}\ro\jumpd{j} \right\}\,, 
 \label{140}
\end{equation}
with the channel operators $\jump{x,y} = \sqrt{\gamma}\A{x,y},$ and
where we have transformed into the interaction picture of $\HO$ which
is given in \ceq{\ref{20}}. For simplicity we have further assumed
that the system is only subject to its free dynamics.

We are interested in the decoherence effects during the time interval
$t \in [t_0,t_0+\tau],$ in between two of the successive projective
measurements which serve to detect quantum jumps (see \fig{fig1}). To
examine these decoherence effects in more detail consider the system
evolution starting from the initially pure state $\ro(t_0) = \rstatee
\lstatee \otimes \rstatev \lstatev,$
where $\rstatee$ denotes the initial electronic state which we include
for completeness here although it remains unaffected by the
dissipative evolution. The initial vibrational state $\rstatev = c_-
\basism + c_+ \basisp$ encodes a qubit in the two basis states
$\basism$ and $\basisp$ which span the subspace
$\subHilbert\subset\Hilbert,$ of the ions motional Hilbert space.
During the time interval $t \in [t_0,t_0+\tau],$ we do not monitor the
dissipative system evolution and the system dynamics suffers motional
decoherence as described by the master equation in
\ceq{\ref{140}}. Given this master equation the time evolved density
operator $\ro(t)$ can be expressed with the help of a Dyson expansion
\cite{louisell73,steinbach95}
\begin{eqnarray}
 \ro(t) &=& \sum_{n=0}^{\infty} \sum_{\{j_1,j_2,...,j_n\}}
	    \int_{t_0}^t dt_n\, ... \int_{t_0}^{t_3} dt_2\, 
            \int_{t_0}^{t_2} dt_1 \nonumber \\
        && \times \left\{\U{t}{t_n} \jump{j_n} ... 
	     \U{t_2}{t_1}\jump{j_1}\U{t_1}{t_0} \right. \nonumber \\
        && \left. \ro(t_0)
             \Ud{t_1}{t_0} \jumpd{j_1} ... \Ud{t}{t_n} \right\} \,,
 \label{180}
\end{eqnarray}
where we sum over all possible sequences of $\{j_n,...,j_1\}$ and
where $j_i \in \{x,y\}$.  In this expression the density operator
$\ro(t)$ is written as a mixture of pure states which result from the
non-Hermitian time evolution
\begin{equation}
 \UU{t} = \exp{\left[-\frac{i}{\hbar} \Heff t \right]} \,,
 \label{200}
\end{equation}
governed by $\Heff = - i \hbar ( \jumpd{x} \jump{x} + \jumpd{y}
\jump{y} )/2,$ and interrupted by quantum jumps through the action of
the operators $\jump{x,y} = \sqrt{\gamma} \A{x,y}.$ We can rewrite
\ceq{\ref{180}} as a sum over sub-ensembles $\ro_{n_x n_y}(t)$ which are
characterized by the number of quantum jumps $n_{x,y}$ that these
sub-ensembles have suffered through the decay channels $\jump{x,y}$ in
the time $t$ without specifying the exact times of the jumps
\cite{steinbach95},
\begin{eqnarray}
 \ro(t) &=& \sum_{n_x,n_y=0}^{\infty} p_{n_x n_y} \ro_{n_x n_y}(t) 
	\nonumber \\
 	&=& \sum_{n_x,n_y=0}^{\infty} p_{n_x n_y} 
	 \rpsi{n_x n_y} \lpsi{n_x n_y}
	\nonumber \\
	&& \otimes\rstatee\lstatee\,.
 \label{220}
\end{eqnarray}
Here the states $\rpsi{n_x n_y}$ are normalized, so that the
quantities $p_{n_x n_y}$ give the probabilities for the state
$\rpsi{n_x n_y}$ in the mixture. To leading order in $\gamma t$ we
have $p_{n_x n_y} \propto (\gamma t)^{n_x+n_y},$
\cite{steinbach95}. We keep the sub-ensemble decomposition as in
\ceq{\ref{220}} and for times $\gamma \tau \ll 1$ we neglect all
$p_{n_x n_y}$ of order $(\gamma \tau)^2$ and higher to obtain
\begin{eqnarray}
 \ro(\tau) &=& p_{0_x 0_y} \ro_{0_x 0_y}(\tau) 
	+ p_{1_x 0_y} \ro_{1_x 0_y}(\tau) 
	\nonumber \\
	&&+ p_{0_x 1_y} \ro_{0_x 1_y}(\tau)\,,
 \label{260}
\end{eqnarray}
where we have let $t_0=0,$ for convenience. 

The mixture in \ceq{\ref{260}} is of the form given in \ceq{\ref{8}}
and serves as our starting point to deduce the stable orthogonal basis
states $\basispm$. The three terms in the mixture have the following
interpretation. The first term corresponds to the system evolving
without any quantum jumps, whereas the second and the third term
describe the dissipative effects of a single quantum jump through the
decay channel $\jump{x}$ and $\jump{y}$, respectively, without
specifying the exact time of the jump. From \ceq{\ref{180}} the three
contributions to the mixture can be further evaluated to give
\begin{eqnarray}
 \sqrt{p_{0_x 0_y}} \rpsiT{0_x 0_y} &=& 
	\UU{\tau} \rstatev\,,
 \label{nojumpstate} \\[2ex]
 \sqrt{p_{1_x 0_y}} \rpsiT{1_x 0_y} &=& 
	\sqrt{\frac{e^{\gamma\tau}-1}{\gamma}}
	\jump{x}\UU{\tau} \rstatev\,,
 \label{xjumpstate} \\[2ex]
 \sqrt{p_{0_x 1_y}} \rpsiT{0_x 1_y} &=& 
	\sqrt{\frac{e^{\gamma\tau}-1}{\gamma}}
	\jump{y}\UU{\tau} \rstatev\,,
 \label{yjumpstate}
\end{eqnarray}
where $\rpsiT{0_x,0_y} \in \SubHilbert{0_x,0_y},$ $\rpsiT{1_x,0_y} \in
\SubHilbert{1_x,0_y},$ $\rpsiT{0_x,1_y} \in \SubHilbert{0_x,1_y},$ and
$\UU{\tau}$ has been defined in \ceq{\ref{200}}. Given that these
three subspaces of the motional Hilbert space are mutually orthogonal,
the mixture in \ceq{\ref{260}} can be projected into one of the three
terms as we will show in Section \ref{detection} below. We now
consider the conditions for the unitary reversibility of the
decoherence effects associated with the three possible outcomes of
this projective measurement as described by Eqs.\,(\ref{nojumpstate})
- (\ref{yjumpstate}).

\subsection{Reversibility of No-Jump Evolution}
\label{NOJUMPS}

First, consider the dissipative system evolution in the absence of
quantum jumps as described by \ceq{\ref{nojumpstate}}. Following
\ceq{\ref{nccond}} the associated decoherence effects are unitarily
reversible if
\begin{equation}
 \lstate \UUd{\tau}\UU{\tau} \rstate =
 \mu^2\,, \qquad \forall \rstate \in \subHilbert\,,
 \label{nojump}
\end{equation}
and where $\subHilbert$ is spanned by $\basispm$. 
It is not possible to find two orthogonal states $\basispm$, which
satisfy this condition for arbitrary times $\tau$, if they are
restricted to a {\em single\/} degree of freedom of the ion's
motion. This becomes clear from the following argument. Let the two
basis states $\basispm = \sum_{n} c_n^\pm \rvib{n},$ where $\rvib{n}$
denote Fock states of the ion's motion along, say, the $x$
direction. For the decoherence effects in the absence of quantum jumps
to be reversible for those basis states we will require from
\ceq{\ref{nojump}}
\begin{displaymath}
 \lbasispm\,e^{-\gamma \Ad{x}\A{x} \tau} \basispm =
 \sum_n\,e^{-\gamma n \tau} |c_n^\pm|^2 =
 \mu^2, 
\end{displaymath}
which implies $|c_n^+| = |c_n^-|, \forall n,$ and the state $\basism$
can be written as $\basism = \sum_{n} c_n^+\,e^{i\phi_n} \rvib{n}.$
Now, for the basis states $\basispm$ to be orthogonal the phases
$\phi_{n}$ cannot all be equal and there exists a pair of phases
$\phi_{n_1} \neq \phi_{n_2}.$ We can then define a superposition state
$\rstate = (\basisp - \,e^{-i\phi_{n_1}}\basism)/\sqrt{2},$ which
obviously satisfies $\rstate \in \subHilbert$ but for which
\ceq{\ref{nojump}} is not satisfied as can be easily shown.
We have performed a numerical search and found orthogonal single-mode states
which satisfy \ceq{\ref{nojump}} for {\em specific\/} times $\tau.$
However, these states depend on the particular value of $\tau$ and
$\gamma$ which limits their applicability and we will not present any
further analysis of such states here.

For the orthogonal basis states to be stable and independent of the period $\tau$
of the projective measurements that serve to detect quantum jumps, they
have to comprise of at least two bosonic modes. To derive such states
in two of the ion's motional degrees of freedom we impose
\begin{equation}
	\frac{i}{\hbar}\Heff \basispm=\Gamma\basispm\,,
	\label{iso}
\end{equation}
on the basis states $\basispm,$ which span $\SubHilbert{0}.$ This
choice satisfies \ceq{\ref{nojump}} and ensures that the dissipative
evolution in the absence of quantum jumps is unitarily reversible. In
fact, \ceq{\ref{iso}} implies 
\begin{equation}
 \UU{t}\rstate = e^{-\Gamma t}\rstate 
	\qquad\forall \rstate \in \SubHilbert{0}\,,
 \label{invariant}
\end{equation}
so that the subspace $\SubHilbert{0}$ remains invariant and
$\SubHilbert{0_x,0_y} = \SubHilbert{0}.$ The general form for states
satisfying \ceq{\ref{iso}} is $  \basispm=\sum_{n=0}^{N} c^\pm_n
\rtwomode{n}{N-n},$ 
where $\Gamma=\gamma N/2$, and where $\rtwomode{n_x}{n_y}$ denote the
usual number state basis for the ion's harmonic motion in the $x$ and
$y$ directions.

\subsection{Reversibility of Quantum Jumps}

To continue our analysis of the stable basis states we consider the
conditions for the unitary reversibility of the decoherence effects
associated with a single quantum jump as described by
Eqs.\,(\ref{xjumpstate}) and (\ref{yjumpstate}). Following
\ceq{\ref{nccond}} these decoherence effects are unitarily reversible
if 
\begin{equation}
 \lstate\jumpd{x,y}\jump{x,y}\rstate = \mu_{x,y}^2 
	\qquad \forall \rstate \in \SubHilbert{0}\,,
 \label{jumpconds}
\end{equation}
where we have used \ceq{\ref{invariant}}. These conditions were
satisfied in the CQED situation of Ref.\cite{inversion} by choosing
$|\psi_+\rangle=|2,0\rangle$, and $|\psi_-\rangle=|0,2\rangle$,
together with the fact that the master equation, as given in
\ceq{\ref{140}}, is form invariant under the general Bogolyubov
transformation
\begin{equation}
 \left(\begin{array}{c}
 	\jump{x} \\ \jump{y} 
 \end{array}\right)=
 \sqrt{\gamma}
 \left[ \begin{array}{cc}
	\cos{\theta} & e^{i\phi}\sin{\theta}\\
	-e^{-i\phi}\sin{\theta} & \cos{\theta} 
 \end{array}\right]
 \left(\begin{array}{c}
	\hat{a}_x \\ \hat{a}_y 
 \end{array}\right)\,,
 	\label{bog}
\end{equation} 
and the operators $\jump{x,y}$ fulfill the conditions of
\ceq{\ref{jumpconds}} for $\theta=\pi/4$. In Ref.\cite{inversion} the
detected quantum jumps were associated with the operators
$\jump{x,y}$, for the values $\theta=\pi/4$ and $\phi=0$. In general,
for the same underlying master equation, the quantum trajectories
formalism yields different unravelings for different forms of the
operators $\jump{x,y}$.  The particular
values of the parameters $\theta$ and $\phi$ are determined by the associated
measurement strategy which monitors the dissipative system
evolution. For $\theta=\pi/4$, these measurements yield no information
on the superposition of the basis states $|\psi_\pm\rangle$, and, so
long as one can physically realize such {\em stabilizing\/}
measurements, one can unitarily invert a quantum jump which occurs though
either the $\jump{x}$ or $\jump{y}$ decay channel. In the scheme of
Ref.\cite{inversion}, the measurement which stabilizes the unknown
quantum superposition against a 
single quantum jump is
accomplished through the use of a beam-splitter \cite{gou96}. In our
case, however, to perform an analogous measurement, that of detecting
the loss of rotational quanta (when $\theta=\pi/4$ and $\phi=\pi/2$),
is difficult. Instead we make use of a duality which exists between
the form of the basis states and the stabilizing measurement. We
choose,
\begin{displaymath}
 \basisp=\rrotmode{0}{2} \,, \qquad
 \basism=\rrotmode{2}{0}\,,
\end{displaymath}
where the states 
\begin{displaymath}
 \rrotmode{n_d}{n_g} \equiv \frac{1}{\sqrt{\footnotesize n_g!n_d!}}\,
 \Ad{d}^{n_d}\Ad{g}^{n_g}\rtwomode{0}{0}\,,
\end{displaymath}
are Fock states of the transformed operators $\A{g} = (\A{x} +
e^{i\phi}\A{y}) / \sqrt{2}$ and $\A{d} = (\A{y}-e^{-i\phi}\A{x})/
\sqrt{2},$ which correspond to the transformation of \ceq{\ref{bog}}
for $\theta=\pi/4.$ In the Cartesian number state basis the states
$\basispm$ are given by
\begin{displaymath}
 \basispm=\frac{1}{2}\left\{ \rtwomode{2}{0} 
	+ e^{-2i\phi}\rtwomode{0}{2}
	\pm \sqrt{2}e^{-i\phi}\rtwomode{1}{1} \right\}\,.
\end{displaymath}
For $\phi=\pi/2$, these are Schwinger two-mode rotational states. One
can show that for $\jump{x,y} = \sqrt{\gamma} \A{x,y}$ the above basis
states $\basispm$ satisfy the conditions given in Eqs.\,(\ref{nojump})
and (\ref{jumpconds}). However, as we have discussed in Section
\ref{strategy}, to accomplish the detection of quantum jumps we
require that the decoherence effects described by Eqs.\,(\ref{nojumpstate})
- (\ref{yjumpstate}), must lead to mutually orthogonal subspaces of the
motional Hilbert space. For the above basis states $\basispm$ this is
not the case since they lead to $\SubHilbert{1_x,0_y} =
\SubHilbert{0_x,1_y} = {\it span} \left\{\rtwomode{1}{0},
\rtwomode{0}{1}\right\}.$ The orthogonality requirement can however be
satisfied by doubling all the excitation numbers in the basis states,
and we define
\begin{equation}
 \sbasispm \equiv \frac{1}{2} \left\{\rtwomode{4}{0} 
	+ e^{i\phi_1}\rtwomode{0}{4}
	\pm \sqrt{2}e^{i\phi_2}\rtwomode{2}{2} \right\}\,,
 \label{mystates}
\end{equation}
where we have given the most general form with non-zero relative
phases $\phi_1$ and $\phi_2$. These states are stable and represent
the main result of this section. They span a subspace $\SubHilbert{0}$
which satisfies the conditions given in Eqs.\,(\ref{nojump}) and
(\ref{jumpconds}), so that the decoherence effects associated with either 
a single jump
or no jump,
are unitarily reversible. Given an initial vibrational state $\rstatev
= c_- \sbasism + c_+ \sbasisp$ which encodes an arbitrary superposition
in the stable basis states $\sbasispm$, we find from
Eqs.\,(\ref{nojumpstate}) - (\ref{yjumpstate})
\begin{eqnarray}
 \rpsiT{0_x 0_y} &=& c_- \sbasism + c_+ \sbasisp\,,
 \label{nojstate} \\
 \rpsiT{1_x 0_y} &=& c_- \xbasism + c_+ \xbasisp\,,
 \label{xjstate} \\
 \rpsiT{0_x 1_y} &=& c_- \ybasism + c_+ \ybasisp\,,
 \label{yjstate}
\end{eqnarray}
and the probabilities in the mixture of \ceq{\ref{260}} can be
evaluated to give $p_{0_x 0_y} = e^{-4 \gamma \tau}$ and $p_{1_x 0_y}
= p_{0_x 1_y} = 2 e^{-4\gamma \tau} (e^{\gamma \tau} - 1).$ 
The decoherence effects lead to mutually orthogonal subspaces which
are given by $\SubHilbert{0_x 0_y} \equiv {\it span} \left\{ \sbasisp,
\sbasism \right\}$, $\SubHilbert{1_x 0_y} \equiv {\it span} \left\{
\xbasisp, \xbasism \right\}$ and $\SubHilbert{0_x 1_y} \equiv {\it
span} \left\{ \ybasisp, \ybasism \right\}$, where we have defined
\begin{eqnarray}
 \xbasispm &=& \frac{1}{\sqrt{2}}\left\{ 
	\rtwomode{3}{0} \pm \rtwomode{1}{2}\right\}\,, 
	\label{Xbasis} \\
 \ybasispm &=& \frac{1}{\sqrt{2}}\left\{
	\rtwomode{0}{3} \pm \rtwomode{2}{1}\right\}\,,
	\label{Ybasis}
\end{eqnarray}
and we have taken $\phi_{1,2}=0$ in the definition of the stable basis
states in \ceq{\ref{mystates}}. Due to the very special properties of
these basis states, the decoherence effects cause the unknown
superposition state $\rstatev = c_- \sbasism + c_+ \sbasisp$ to evolve
into a mixture of superpositions with identical coefficients $c_\pm$,
but in orthogonal subspaces. If, through the proper measurement, we
learn {\em only\/} which Hilbert subspace the system resides, then
that subspace possesses an exact copy of the original qubit state and
through a unitary manipulation we can restore this information to
${\cal H}_0$. This does not contradict the no-cloning theorem
\cite{wootters82}, since once we have made the measurement, the
``copies'' of the qubit in the other Hilbert subspaces are lost. 

We emphasize that the basis states derived in \ceq{\ref{mystates}} are
stable in the presence of a finite-temperature reservoir even though
for their derivation it was sufficient to consider the
zero-temperature case. It is straightforward to show that the subspace
$\SubHilbert{0} \equiv {\it span}\left\{\sbasisp, \sbasism \right\}$,
also satisfies the conditions given in Eqs.\,(\ref{nojump}) and
(\ref{jumpconds}) for the finite-temperature case where the associated
decay channel operators are given by \ceq{\ref{60}}. In this context
we note that recent work on the more formal aspects of states of
bosonic systems that can be stabilized in the presence of a
zero-temperature reservoir has also uncovered the states given in
\ceq{\ref{mystates}} and many more complicated multi-mode states
\cite{Chuang}. In addition to the results reported in
Ref.\cite{Chuang} the analysis presented here demonstrates the
potential of those states for the more general, finite-temperature
case. In the remainder of this paper we concentrate on the case of a
zero-temperature reservoir as we have already noted above.

\section{Detection of decoherence processes}
\label{detection}

In the following we discuss how to detect the decoherence processes
associated with the master equation given in \ceq{\ref{140}}. More
specifically, based on our specific choice of encoding information in
the motional basis states derived in \ceq{\ref{mystates}}, we present
a method that determines whether a quantum jump associated with the
decay channel $\jump{x}$ or $\jump{y}$ has occurred. The corresponding
processes constitute step (i) in our stabilisation scheme as depicted
in \fig{fig1}.

The detection of a quantum jump is accomplished in two steps. The
first step serves to generate an ``error syndrome'' which only carries
information about the effects of decoherence. To be more specific,
given that the initial electronic state of the ion $\rstatee =
\rground,$ we show in Section \ref{entanglement} how to entangle the
mixed state that results from the decoherence processes, as in
\ceq{\ref{260}}, with the electronic states $\rground$ and
$\rexcited$. In essence, we generate a unitary operator
$\hat{U}_{ent}$, which, in its generic form, is described by
\begin{eqnarray}
 \hat{U}_{\it ent}\rstatev \otimes \rground 
 &=& \rstatev \otimes \rexcited \,,
	\quad \forall \rstatev \in \SubHilbert{\it ent}\,,
	\nonumber \\
 \hat{U}_{\it ent} \rstatev \otimes \rground
 &=& \rstatev \otimes \rground \,,
	\quad \forall \rstatev \in \SubHilbert{\it ent}^{\perp}\,,
	\nonumber \\
 && \label{mapping}
\end{eqnarray}
where $\SubHilbert{\it ent}$ is a specific subspace of the ion's
motional Hilbert space and $\SubHilbert{\it ent}^{\perp}$ is the
orthocomplement of $\SubHilbert{\it ent}$. It is important that this
operation leaves the motional state $\rstatev$ unaffected. Also it is
necessary that the operator $\hat{U}_{\it ent}$ entangles the motional
subspaces $\SubHilbert{\it ent}^{\perp}$ and $\SubHilbert{\it ent}$
with the electronic states $\rground$ and $\rexcited$ {\it without}
supplying any further information on the vibrational state $\rstatev.$
This can be accomplished by making use of our {\em a priori\/}
knowledge of the motional basis states and the decoherence processes
discussed in Section \ref{dissipation}. 
The above entanglement process produces a binary signature in form of
the ion's electronic state which serves as our error syndrome. The
second step of our quantum jump detection scheme serves to ``read
out'' this error syndrome. This is accomplished through a projective
measurement of the ion's electronic state which we discuss in Section
\ref{projection}. Following the entanglement operation given in
\ceq{\ref{mapping}}, the detection of the ion in the electronic state
$\rground$ ($\rexcited$) amounts to projecting the motional state of
the ion onto the subspace $\SubHilbert{\it ent}^{\perp}$
($\SubHilbert{\it ent}$).

Before we describe in detail how the error syndrome can be generated
and read out we briefly elaborate further on the sequence of
operations which constitute step (i) in \fig{fig1}. First, the
detection of a quantum jump associated with the decay channel
$\jump{x}$ is accomplished through constructing the entanglement
operation such that $\SubHilbert{\it ent} = \SubHilbert{1_x 0_y}$ in
\ceq{\ref{mapping}}. A subsequent measurement of the ion in the
electronic state $\rstatee = \rexcited$ is then equivalent to the
detection of a quantum jump as described by \ceq{\ref{xjumpstate}}, and
the motion of the ion is projected into the pure state given in
\ceq{\ref{xjstate}}. If, on the other hand, the projective measurement
yields the result $\rstatee = \rground$, for the electronic state of
the ion, {\em no\/} quantum jump associated with the decay channel
$\jump{x}$ has occurred and the ion is left in a mixture of the states
given in Eqs.\,(\ref{nojstate}) and (\ref{yjstate}). Irrespective of
the outcome of this first projective measurement we then determine
whether or not a quantum jump associated with the decay channel
$\jump{y}$ has occured. The corresponding procedure is identical to
the one which serves to detect a quantum jump through the decay
channel $\jump{x}$ but where now $\SubHilbert{\it ent} =
\SubHilbert{0_x 1_y}$ for generating the error syndrome. If through
the first measurement we have detected a quantum jump associated with
the decay channel $\jump{x}$ the outcome of the second measurement
will always be $\rstatee = \rground$ and we do not obtain further
information. However, for the case where no quantum jump through the
decay channel $\jump{x}$ has occured, the outcome $\rstatee =
\rexcited$ for the second projective measurement corresponds to the
detection of a quantum jump as described by \ceq{\ref{yjumpstate}} and
projects the ion into the pure motional state given in
\ceq{\ref{yjstate}}. On the other hand the measurement of the ion in
the electronic state $\rstatee = \rground$ for a second time implies
that no quantum jump has occurred through either of the decay channels
and the ion is projected into the pure motional state given in
\ceq{\ref{nojstate}}. The second projective measurement thereby
completes the sequence of operations that constitute step (i) in
\fig{fig1} and leaves the ion in a definite, pure motional state. From
this the original motional state can be restored in step (ii) of our
stabilisation scheme as indicated in \fig{fig1} and as we describe in
Section \ref{repair}. In the following we now detail the processes
which accomplish the generation and reading out of the error syndrome.

\subsection{Entanglement Stage}
\label{entanglement}

The physical process that we use to generate the essential part of the
entanglement operation given in \ceq{\ref{mapping}} is a resonant
stimulated Raman transition between the two ground states $\rground$
and $\rexcited$ in a specific laser geometry. This generates Rabi
oscillations between the states $\rground$ and $\rexcited$ where the
Rabi frequency depends on the ion's motional state. An essential
requirement for the generation of the desired entanglement is the
accurate control of the phases acquired through these Rabi
oscillations. We therefore consider time dependent laser pulses
characterized through their dimensionless pulse shape $f(t)$ to
generate the Raman transitions, and the phase of the Rabi oscillations
is controlled through the generalized pulse area
\begin{equation}
 A = \int_{-\infty}^{\infty} f(t^\prime)^2 dt^\prime\,.
 \label{395}
\end{equation} 
For the generation of the error syndrome associated with the decay
channels $\jump{x}$ and $\jump{y}$, we will be interested in the case
where the Raman-induced Rabi oscillations are only sensitive to the
ion's motion along one of the principal axes. As shown in Appendix
\ref{ramanapp}, this can be realized by resonantly exciting the
two-photon transition $\rground \Leftrightarrow \rexcited$ with two
plane waves having wavevectors $\k{\ground}$ and $\k{\excited},$ and
where the wavevector difference $\underline{\delta k} = \k{\ground} -
\k{\excited}$ is aligned with that axis. The resulting Hamiltonian is
given by
\begin{eqnarray}
 \H &=& - \hbar \gge(t)
	 \exp{\left[-\ETA{j}^2/2\right]} \,\sm \nonumber \\
    && \otimes \sum_{n=0}^\infty 
	 \frac{(-i\ETA{j})^{2 n}}{n! n!} \Ad{j}^{n} \A{j}^{n} 
	+ {\rm H.c.}\,,
 \label{HRaman}
\end{eqnarray}
where $j=x$ for the case of aligning $\underline{\delta k}$ with the
$x$ axis, and $j=y$ for the case of aligning $\underline{\delta k}$
with the $y$ axis. The symbols $\ETA{x}$ and $\ETA{y}$ denote the
Lamb-Dicke parameters
\begin{equation}
  \ETA{x} = \Delta x_0\, \left|\underline{\delta k}\right|\,,\qquad
  \ETA{y} = \Delta y_0\, \left|\underline{\delta k}\right|\,,
  \label{ldp}
\end{equation}
for the case $j=x,y,$ as explained above, and $\dx = (\hbar / 2 \Nu{x}
m)^{1/2}$ and $\dy = (\hbar / 2 \Nu{y} m)^{1/2}$ give the widths of
the motional ground state along the $x$ and the $y$ direction,
respectively. The Raman coupling constant $\gge(t) \propto f(t)^2$ and
its exact form is given in \ceq{\ref{411}}.
To generate the required entanglement operations we assume the
Lamb-Dicke limit \cite{Lamb-Dicke,blockley92}, which can be defined
through $\lstatev \delta k_x^2 \x^2 \rstatev \ll 1,$ and $\lstatev
\delta k_y^2 \y^2 \rstatev \ll 1,$ 
and which implies $\ETA{x,y} \ll 1,$ for the Lamb-Dicke
parameters. This allows us to retain only the lowest order terms in
$\ETA{x,y}$ in the Hamiltonian describing the Raman-induced dynamics.

We first address the question of detecting whether a quantum jump
associated with the operator $\jump{x}=\sqrt{\gamma}\A{x}$ has
occurred. The corresponding error syndrome is generated through an
entanglement operation
as in \ceq{\ref{mapping}} with $\SubHilbert{\it ent} = \SubHilbert{1_x
0_y},$ and $\SubHilbert{\it ent}^{\perp} = \SubHilbert{0_x 0_y} \oplus
\SubHilbert{0_x 1_y},$
which can be accomplished by exploiting the fact that for all states
$\rstatev \in \SubHilbert{1_x 0_y}$ the motional excitation numbers
$n_x$ along the $x$ direction are {\em odd\/}, whereas for states
$\rstatev \in \SubHilbert{0_x 0_y} \oplus \SubHilbert{0_x 1_y}$ the
motional excitation numbers $n_x$ are {\em even\/}, as seen from
Eqs.\,(\ref{mystates}), (\ref{Xbasis}) and (\ref{Ybasis}). With this
in mind, we construct a laser exciting field where the induced
stimulated Raman transition is only sensitive to the ion's motion
along the $x$ direction. More specifically, we realize the essential
part of the above entanglement operation through the unitary time
evolution generated by the Hamiltonian
\begin{equation}
 \hat{H}^{x} = \hbar g(t) \Ad{x}\A{x}
 \otimes \rground \lexcited + {\rm H.c.}\,,
 \label{hamI}
\end{equation}
for a specific choice of the generalized pulse area given in
\ceq{\ref{395}}. This form of Hamiltonian has previously been
discussed as a degenerate Raman-coupled model in the context of CQED
\cite{knight86}. More recently, Gerry has shown how such an
interaction may be realized for a trapped ion by driving a dipole
transition in a specific laser arrangement \cite{gerry97}.

\noindent
\begin{minipage}{0.48\textwidth}
\begin{figure}[h]
   {
   %
    \leavevmode
    \epsfxsize=80mm
    \epsffile{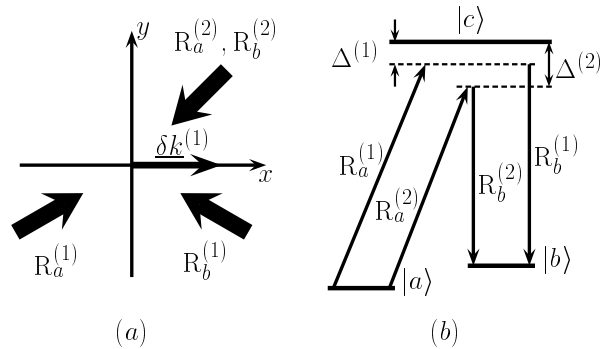}
   %
   }
\caption{Laser arrangement which generates the error syndrome for a
quantum jump associated with the decay channel $\jump{x}$. (a) Two
pairs of lasers drive a stimulated Raman transition between the states
$\rground$ and $\rexcited,$ which is sensitive to the ion's motion
along the $x$ direction. The first pair, ${\rm
R}_{\ground,\excited}^{(1)}$ is arranged so that the wavevector
difference $\underline{\delta k}^{(1)} = \k{\ground}^{(1)} -
\k{\excited}^{(1)},$ is aligned with the $x$ axis. The second pair
${\rm R}_{\ground,\excited}^{(2)}$ is co-propagating, so that
$\underline{\delta k}^{(2)} = \k{\ground}^{(2)} - \k{\excited}^{(2)} =
\underline{0}.$ (b) The stimulated absorption and emission processes
induced by the laser excitation can be treated separately for the two
pairs of Raman lasers, if the detunings $\Delta^{(1)}$ and
$\Delta^{(2)}$ are chosen sufficiently different.
}
\label{fig3}
\end{figure}
\vspace{1pt}
\end{minipage}

For the dipole-forbidden $\rground \Leftrightarrow \rexcited$
transition which is relevant to us here, the form of coupling given in
\ceq{\ref{hamI}} can be generated through a resonant stimulated Raman
transition between the electronic states $\rground$ and $\rexcited$
which is induced by two pairs of laser beams as shown in \fig{fig3}. 
The first pair of Raman lasers is arranged so that the wavevector
difference $\underline{\delta k}^{(1)} = \k{\ground}^{(1)} -
\k{\excited}^{(1)},$ is aligned with the $x$ axis. Following
\ceq{\ref{HRaman}} this generates the Hamiltonian
\begin{eqnarray}
 \H^{(1)} &=& - \hbar \gge^{(1)}(t)
	 \exp{\left[-\eta^2/2 \right]}
	 \,\sm \nonumber \\
    && \otimes \left\{ {\bf 1} - \eta^2 \Ad{x}\A{x} \right\}
	+ {\rm H.c.}\,.
 \label{hamIa}
\end{eqnarray}
where we have assumed the Lamb-Dicke limit \cite{Lamb-Dicke}, and kept
only the leading order terms in the Lamb-Dicke parameter $\eta =
\ETA{x}^{(1)}.$ The second pair of Raman lasers serves to cancel the
zeroth-order contribution in this Hamiltonian which is insensitive to
the motional state of the ion. The laser geometry is arranged such
that the wavevector difference $\underline{\delta k}^{(2)} =
\k{\ground}^{(2)} - \k{\excited}^{(2)} = \underline{0},$ vanishes
exactly
\cite{Turchette}. From \ceq{\ref{HRaman}} this gives rise to the
Hamiltonian
\begin{equation}
 \H^{(2)} = - \hbar g^{(2)}(t) \sm + {\rm H.c.} \,,
 \label{hamIb}
\end{equation}
since $\ETA{x}^{(2)} = \ETA{y}^{(2)} = 0.$ The combination of the two
pairs of Raman lasers is described by the Hamiltonian $\H = \H^{(1)} +
\H^{(2)},$ if the stimulated absorption and emission processes induced
by the laser excitations can be treated separately for the two pairs of
Raman beams. This is the case if the detunings $\Delta^{(1)}$ and
$\Delta^{(2)}$ of the lasers from the excited state $\rraman$
(\ceq{\ref{ramandet}}) are chosen sufficiently different for the two
pairs of Raman beams as depicted in \fig{fig3}(b). In this situation, and
if we assume the laser phases and amplitudes to be arranged such that
$g^{(2)}(t) = - \exp{[-\eta^2/2]} g^{(1)}(t),$ the laser exciting
field described above generates the Hamiltonian given in
\ceq{\ref{hamI}} and the coupling constant
\begin{equation}
 g(t) = \eta^2 \exp{[-\eta^2 / 2]} g^{(1)}(t) = g f(t)^2\,.
 \label{couplingconst}
\end{equation}
For the Hamiltonian in \ceq{\ref{hamI}} one can calculate the unitary
operator which describes the resulting time evolution to be
\begin{eqnarray}
 \hat{U}(A)^{x} &=& 
 \cos{\left[\hat{\chi} A\right]} \left(\rground\lground 
			+ \rexcited\lexcited\right)
	\nonumber\\
 && + \ \sin{\left[\hat{\chi} A\right]} \left(\rground\lexcited 
				- \rexcited\lground\right) \,,
	\label{unitaryI}
\end{eqnarray}
where $\hat{\chi} = |g| \Ad{x}\A{x}$, and we have further assumed
$g=i|g|$ for the phase of the coupling constant. The generalized pulse
area $A$ is given by \ceq{\ref{395}}. If we now choose $A =
\pi/2|g|,$ we obtain the mapping
\begin{equation}
 \hat{U}(A)^{x}\xbasispm \otimes \rground = \xbasismp \otimes
 \rexcited\,, \label{xmapping}
\end{equation}
for the basis states $\xbasispm$ which are given in \ceq{\ref{Xbasis}}
and which span the motional subspace $\SubHilbert{1_x 0_y}$ 
associated with the decay channel $\jump{x}.$ On the other hand, for
the basis states $\sbasispm$ and $\ybasispm$ which are given in
Eqs.\,(\ref{mystates}) and (\ref{Ybasis}), and which span the motional
subspace $\SubHilbert{0_x 0_y} \oplus \SubHilbert{0_x 1_y},$ the time
evolution generates the mapping
\begin{equation}
 \hat{U}(A)^{x}\basispm_{0,y} \otimes \rground = \basismp_{0,y} \otimes \rground\,,
 \label{othermapping}
\end{equation}
so that the laser arrangement described above generates the essential
feature of the entanglement operation given in \ceq{\ref{mapping}}. It
transfers the binary information as to whether or not a quantum jump
associated with the decay channel $\jump{x}$ has occurred, into the
electronic degrees of freedom. However, the Raman-induced operation
does not leave the motional state of the ion unaffected. As seen from
Eqs.\,(\ref{xmapping}) and (\ref{othermapping}), it interchanges the
basis states that encode the quantum information in the subspaces
$\SubHilbert{0_x 0_y}$, $\SubHilbert{1_x 0_y}$ and $\SubHilbert{0_x
1_y}.$ However, after determining whether or not a quantum jump has
occurred through the $\jump{x}$ channel, we must repeat the
interrogation to determine whether or not a jump has occurred through
the $\jump{y}$ channel. This second interrogation again interchanges
the motional basis states and thus after both steps are completed, the
motional state is returned to its original configuration.

\noindent
\begin{minipage}{0.48\textwidth}
\begin{figure}[h]
   {
   %
    \leavevmode
    \epsfxsize=80mm
    \epsffile{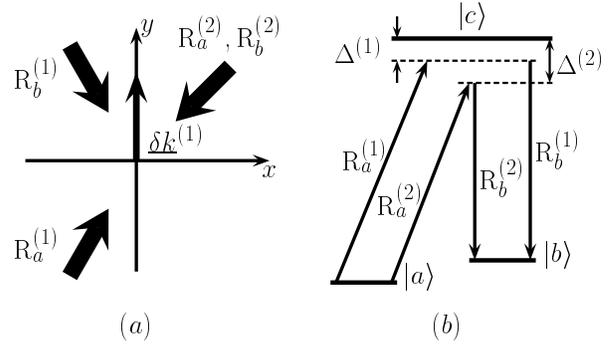}
   %
   }
\caption{Laser arrangement which generates the error syndrome for a
quantum jump associated with the decay channel $\jump{y}$. (a) The
laser geometry is almost identical to the one depicted in \fig{fig3}
but where the first pair ${\rm R}_{\ground,\excited}^{(1)}$ is now
arranged so that the wavevector difference $\underline{\delta k}^{(1)}
= \k{\ground}^{(1)} - \k{\excited}^{(1)},$ is aligned with the $y$
axis. (b) The laser frequencies remain the same as in generating the
error syndrome for the $\jump{x}$ decay channel.}
\label{fig4}
\end{figure}
\vspace{1pt}
\end{minipage}

The detection of the quantum jump associated with the decay channel
$\jump{y}$ is achieved through an almost identical procedure. The
corresponding error syndrome is generated through the entanglement
operation
\begin{eqnarray}
 \hat{U}(A)^{y}\basispm_{0,x} \otimes \rground &=& \basismp_{0,x} \otimes \rground\,,
	\nonumber \\
 \hat{U}(A)^{y}\ybasispm \otimes \rground &=& \ybasismp \otimes \rexcited\,.
\label{entII}
\end{eqnarray}
This mapping is generated in an almost identical manner as the one
which serves to generate the error syndrome for the decay channel
$\jump{x}.$
The unitary operator $\hat{U}(A)^y$ can be realized through resonantly
driving the $\rground \Leftrightarrow
\rexcited$ transition with two pairs of Raman lasers as described
above but in a slightly different laser geometry. This is illustrated
in \fig{fig4}. The first pair of lasers is now arranged so that the
wavevector difference $\underline{\delta k}^{(1)} = \k{\ground}^{(1)}
- \k{\excited}^{(1)},$ is aligned with the $y$ axis, and the resulting
Raman transition is only sensitive to the motion of the ion along the
$y$ axis. The geometry of the second pair of Raman lasers and the
arrangement of the laser phases and amplitudes remain the same, so
that with the replacement $x \rightarrow y$ the analysis is identical
to the discussion that led us from \ceq{\ref{hamIa}} to
\ceq{\ref{unitaryI}}. The resulting time evolution is described by
\begin{eqnarray}
 \hat{U}(A)^{y} &=& 
 \cos{\left[\hat{\chi} A\right]} \left(\rground\lground 
			+ \rexcited\lexcited\right)
	\nonumber\\
 && + \ \sin{\left[\hat{\chi} A\right]} \left(\rground\lexcited 
				- \rexcited\lground\right) \,,
	\label{unitaryII}
\end{eqnarray}
where $\hat{\chi} = |g| \Ad{y}\A{y}$, and the coupling constant $g$ is
defined in \ceq{\ref{couplingconst}}. With the choice $A =
\pi/2|g|,$ the time evolution generates the mapping given in
\ceq{\ref{entII}} and the error syndrome for the detection of quantum
jumps associated with the decay channel $\jump{y}$ can be read out
through a measurement of the ion's electronic states $\rground$ and
$\rexcited$ as we will describe in Section \ref{projection} below.

To give an estimate of the time scales involved in the entanglement
stage discussed here, we consider the specific example of $\rm Be^+$
ions and associate the electronic states $\rground$, $\rexcited$ and
$\rraman$ as indicated in \fig{fig2}(b). This is the level scheme that
was employed by Monroe {\em et al.\/} \cite{ionqc}, for the first
demonstration of quantum logic with trapped ions and in further
experiments in the group of Wineland at NIST
\cite{wineland98,meekhof96}. To remain specific we model the explicit
time dependence of the laser pulses through the dimensionless pulse
shape
\begin{equation}
 f(t) = \sin^2{\left[\frac{\pi t}{\Tl}\right]}\,,
 \qquad 0 \le t \le \Tl\,,
 \label{pulseshape}
\end{equation}
where $\Tl$ denotes the pulse duration. We can then evaluate the
generalized pulse area in \ceq{\ref{395}} to obtain $A = 3 \Tl/8.$ The
generation of the error syndrome through the entanglement operations
in Eqs.\,(\ref{xmapping}) - (\ref{entII}) requires $A = \pi/2|g|,$ for
the pulse area which implies $\Tl = 4 \pi/ 3|g|,$ for the pulse
duration. To estimate this time we calculate the value of $|g|$ from
\ceq{\ref{couplingconst}}. We assume $|g^{(1)}|/2\pi = 500 \rm kHz$
for the peak value of the coupling constant generated by the first
pair of Raman lasers as in \ceq{\ref{hamIa}} and $\eta = 0.2,$ for the
corresponding Lamb-Dicke parameter. These are parameters taken from
the experiment reported in \cite{meekhof96}. From
\ceq{\ref{couplingconst}} we find $|g|/2\pi = 20 \rm kHz,$ and obtain
\begin{equation}
 \Tl \approx 33 \mu s\,,
 \label{limit1}
\end{equation}
for the duration of the pulse which generates
the error syndrome. This can be shortened through an increase in the
value of $|g|$. However, to remain within the validity of our analysis
we have to consider the limit set by the vibrational rotating wave
approximation that was made in Appendix \ref{ramanapp} in deriving the
Hamiltonian in \ceq{\ref{HRaman}}. This limit is given by the
off-resonant excitation of the first vibrational sideband through the
first pair of Raman lasers and requires $\eta |g^{(1)}| / \Nu{x,y} \ll
1,$ \cite{Ciracadv}. For the quoted experimental parameters and
$\Nu{x,y}/2\pi = 10 \rm MHz,$ \cite{meekhof96}, we find $\eta
|g^{(1)}| / \Nu{x,y} = 10^{-2},$ so that the duration of the entanglement
operation may be shortened by one order of magnitude. To go further
would however require an increase in the trap frequency to remain
within the low-excitation regime.


It is important to contrast the entanglement operation described here
with the transfer of a quantum state between atomic ground-state
Zeeman sublevels and bosonic degrees of freedom as described by
Parkins {\em et al.\/} through adiabatic processes in a CQED situation
\cite{Parkins}. There, a {\em quantum state\/} is transferred, whereas
the entanglement operation described here leaves the vibrational state
$\rstatev$ unchanged and transfers {\em information\/} as to whether
this states lies in a specific subspace of the motional Hilbert space.

\subsection{Projective Measurement}
\label{projection}

We now turn to the second step of our quantum jump detection scheme
which reads out the error syndrome generated through the entanglement
operation discussed above. As we have already mentioned, this second
step consists of a projective measurement of the ion's electronic
state. This measurement must not significantly disturb the motional
state of the ion which we aim to stabilise here. Existing measurement
procedures such as atomic shelving and resonance fluorescence
\cite{fluorescence,beige96} entail the emission of a large number of
spontaneous photons in the typical experimental situation where
detector efficiency is less than unity and the solid angle covered is
not the full $4\pi$. In these measurement strategies the center of
mass of the ion receives numerous spontaneous emission recoil kicks in
random directions which results in an overall heating of the ion's
motion. To avoid this we employ a technique which has been discussed
in the literature for the generation of single photon wavepackets
\cite{Gheri,law97}. We adopt here the scheme of Gheri {\em et al.\/}
\cite{Gheri}, where a $\Lambda$-type three-level atom (in our case the
ion whose internal level structure is shown in \fig{fig2}) is
surrounded by a low-Q optical cavity and excited by a time-dependent
laser pulse. The laser and cavity electric fields couple the two
ground states $\rground$ and $\rexcited$ through a resonant two-photon
process and induce a stimulated Raman transition between those
states. The excited state $\rraman$ is detuned by a large enough
amount $\Delta$ that it is never appreciably excited and spontaneous
emission events can be faithfully neglected. The situation is
illustrated in \fig{fig5}. We show below how, with suitable
approximations, the ion's electronic and motional degrees of freedom
decouple so that the ion-cavity system reduces to the model considered
by Gheri {\em et al.\/}. We will not go into the detailed solutions to
the dissipative dynamics since these are discussed in
Ref.\cite{Gheri}. Gheri {\em et al.\/} have shown that if the initial
electronic state $\rstatee = \rexcited$, then the laser pulse
transfers the electronic population into state $\rground$ and
simultaneously excites the cavity mode which rapidly decays into the
external field in the form of a single photon wavepacket. More
specifically, in the bad cavity regime where $\kappa \gg |g(t)|$, they
derive an explicit solution to the dynamics of the ion-cavity system
\cite{Gheri}. Using this solution the probability $P(t)$ for a single
photon to be transmitted from the cavity during the time interval
$[0,t]$ can be calculated as shown in Ref.\cite{law97} and we obtain
\begin{equation}
 P(t) = 1 - \exp \left[ - 2
	\int_0^t \frac{|g(\tp)|^2}{\kappa} d\tp \right]\,,
\label{mygun1}
\end{equation}
where $\kappa$ is the decay rate of the cavity field amplitude, and
$g(t)$ denotes the Raman coupling constant for the transition between
the two ground states $\rground$ and $\rexcited$. Note that $g(t)
\propto \Elaser(t)$, where $\Elaser(t) = f(t) \Elaser$ denotes the
electric field amplitude of the external laser pulse and determines
the time dependence of $g(t)$ through the pulse envelope $f(t)$. 

\noindent
\begin{minipage}{0.48\textwidth}
\begin{figure}
   {
   %
    \leavevmode
    \epsfxsize=80mm
    \epsffile{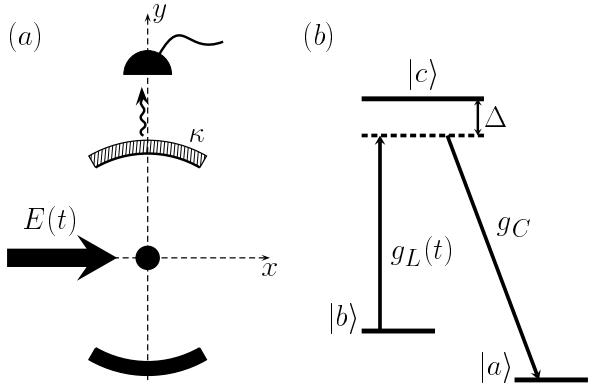}
   %
   }
\caption{To perform a measurement of the ion's electronic states
$\rground$ and $\rexcited$ without significantly disturbing the
motional degrees of freedom the ion is surrounded by a low-Q optical
cavity and excited by a laser pulse with electric field envelope
$E(t)$ as indicated in (a). The laser and cavity electric fields
resonantly couple the two states $\rground$ and $\rexcited$ and induce
a stimulated Raman transition between those states as shown in (b). In
the bad cavity regime this transition is overdamped. The excited state
$\rraman$ is detuned by a large enough amount $\Delta$ that it is
never appreciably excited. If the electronic state of the ion
$\rstatee = \rexcited$, then the laser pulse transfers the population
into state $\rground$ and simultaneously excites the cavity mode which
rapidly decays into the external field in the form of a single photon
wavepacket. On the other hand if the electronic state of the ion
$\rstatee = \rground$, no photon is generated. In this situation the
presence (absence) of a transmitted photon which is observed through
placing a detector in the cavity output channel as indicated in (a),
constitutes a measurement of the ion in its electronic state
$\rexcited$ ($\rground$).}
\label{fig5}
\end{figure}
\vspace{1pt}
\end{minipage}

For a sufficiently large pulse area, the exponential term in
\ceq{\ref{mygun1}} becomes negligible and $P(t) \approx 1.$ This
implies that if the initial state of the ion $\rstatee = \rexcited,$
then the laser pulse triggers the transmission of a single-photon
wavepacket from the cavity with almost unit probability. In this
situation the detection of a transmitted photon through placing a
detector in the cavity output channel as indicated in \fig{fig5}(a)
constitutes, to a very good approximation, a measurement
of the ion in its electronic state $\rexcited$. On the other hand,
if the initial electronic state $\rstatee =
\rground$, no photon is generated since the frequencies of the cavity
and laser electric fields are assumed to be significantly different
from each other, so that they only generate transitions associated
with their own channels as shown in \fig{fig5}(b). Therefore, the
absence of a detection event represents a measurement of the ion in
its electronic state $\rground$. Note that irrespective of the
measurement outcome the ion is left in its electronic state $\rstatee
= \rground$ after the measurement.

In the following we show how the ion-cavity system depicted in
\fig{fig5} reduces, in the low-excitation regime \cite{Ciracadv}
and the Lamb-Dicke limit \cite{Lamb-Dicke,blockley92}, to the model
considered by Gheri {\em et al.\/} \cite{Gheri}. The Hamiltonian which
describes the free evolution of the electronic, motional and cavity
degrees of freedom is given by
\begin{eqnarray}
 \HO &=& \sum_{\j=\ground,\excited,\raman} \hbar \omj \rj \lj 
 + \sum_{\j=x,y} \hbar \Nu{\j} 
	\left(\frac{1}{2} + \Ad{\j} \A{\j}\right)
	\nonumber \\
 && + \ \hbar \omcav \Cd\C \,, 
 \label{mygun2}
\end{eqnarray}
where the operators $\C$ ($\Cd$) are the bosonic annihilation
(creation) operators for the single-mode cavity field excitations and
$\omcav$ denotes the cavity-mode frequency. As depicted in
\fig{fig5}(b) we are considering a {\em resonant\/} two-photon
transition between the states $\rground$ and $\rexcited$ and assume
the triggering laser pulse and the cavity mode to be detuned by the
same amount $\Delta$ from the excited state $\rraman$. The
corresponding electric fields are given by
\begin{eqnarray}
 \Elaser(\x,t) &=& \Elaseramp(t) 
	\,e^{-i[\klaser \x - \omlaser t]}
	+ {\rm H.c.} \,, 
 	\nonumber \\
 \Ecav(\y) &=& \Ecavamp \cos{[\kcav \y]} + {\rm H.c.}\,,
 	\label{mygun3}
\end{eqnarray}
where $\klaser$ denotes the wavevector of the laser and $\kcav$
describes the mode function of the cavity. As seen from the expression
for the cavity electric field, we assume the ion to be positioned at
an antinode of the cavity field mode to minimize the coupling between
the ion's electronic and motional degrees of freedom which arises from
the electric field gradients \cite{wineland98}. In the rotating frame
defined by the Hamiltonian $\H = \HO -\hbar \Delta \rraman\lraman,$
where $\HO$ is given in \ceq{\ref{mygun2}}, the coherent dynamics of the
system are described by the Hamiltonian
\begin{eqnarray}
 \HI(t) &=& \hbar \Delta \rraman \lraman 
	- \hbar \glaser(t)\,\rexcited\lraman\,e^{-i[\klaser \x(t)]} + {\rm H.c.}
 	\nonumber \\
 & & - \hbar \gcav\,\rground\lraman\,\Cd\,
	\cos{[\kcav \y(t)]} + {\rm H.c.}\,,
	\label{mygun4}
\end{eqnarray}
where the coupling constants $\glaser(t) = \lexcited \hat{\wp} \rraman
\Elaseramp(t)/\hbar$, and $\gcav = \lground \hat{\wp} \rraman \Ecavamp
/\hbar$, in terms of the laser pulse amplitude $\Elaseramp(t)$ and the
electric field per photon $\Ecav$ inside the cavity. The symbol
$\hat{\wp}$ denotes the dipole operator and we have made the optical
rotating wave approximation in the above. Note that in the rotating
frame the position operators are explicitely time dependent and are
given as $\x(t) = \dx (\A{x} e^{-i\Nu{x}t} + \Ad{x} e^{i\Nu{x}t}),$
and $\y(t) = \dy (\A{y} e^{-i\Nu{y}t} + \Ad{y} e^{i\Nu{y}t}),$ where
$\dx = ( \hbar/2\Nu{x} m )^{1/2}$ and $\dy = ( \hbar/2\Nu{y} m
)^{1/2}$, give the widths of the motional ground state along the $x$
and $y$ directions, respectively. In the limit of large detuning,
where $|\Delta| \gg |\glaser(t)|, |\gcav|,$ we can adiabatically
eliminate the excited state $\rraman$ and obtain
\begin{eqnarray}
 \HI(t) &=& -\hbar \delta_\ground\,\rground \lground\,
	\Cd\C\,\cos^2{[\kcav \y(t)]}
	-\hbar \delta_\excited(t)\,\rexcited\lexcited
 \nonumber \\
 && - \hbar g(t) \,\rground\lexcited\,\Cd\,
	\cos{[\kcav \y(t)]}\,e^{i[\klaser \x(t)]}
	+ {\rm H.c.}\,,
 \nonumber \\
 &&	\label{mygun5}
\end{eqnarray}
as shown in Ref.\cite{steinbach97}. The first two terms in this
expression correspond to optical Stark shifts arising from the
adiabatic elimination of the excited state and we have introduced the
abbreviations $\delta_\ground = |\gcav|^2/\Delta$ and
$\delta_\excited(t) = |\glaser(t)|^2/\Delta$. The last two terms
describe the stimulated Raman transition between the two ground states
$\rground$ and $\rexcited$, and $g(t) = \gcav\glaser^*(t)/\Delta$
denotes the corresponding Raman coupling constant. To demonstrate how
the dynamics described by the Hamiltonian in \ceq{\ref{mygun5}} reduce
to that of the model of Gheri {\em et al.} \cite{Gheri}, we first
perform the vibrational rotating wave approximation
\cite{blockley92}. To this end we expand the position operators in
\ceq{\ref{mygun5}} in terms of the vibrational creation and
annihilation operators $\Ad{x,y}$ and $\A{x,y}$ in normal-ordered form
\cite{nonlinearJCM}. We then assume the low excitation regime, where
$\delta_\ground, \delta_\excited(t) \ll \Nu{x,y},$ and the Stark
shifts produced by the laser and cavity electric fields are small when
compared with the vibrational energy level spacing
\cite{Ciracadv}. This implies that the resonance conditions are not
significantly modified. Keeping only resonant terms in the Hamiltonian
in \ceq{\ref{mygun5}} we obtain
\end{multicols}
\noindent
\rule{0.5\textwidth}{0.4pt}\rule{0.4pt}{\baselineskip}
\begin{eqnarray}
 \HI(t) &=& -\hbar \delta_\ground\,\rground \lground\,
	\Cd\C\, \otimes \left\{ 1 + e^{-2\ETA{C}^2} \sum_{n=0}^\infty
	\frac{(2 i \ETA{C})^{2n}}{n! n!} \Ad{y}^n \A{y}^n \right\}
	-\hbar \delta_\excited(t)\,\rexcited\lexcited
   	\nonumber \\[2ex]
 && - \hbar g(t)\,\rground\lexcited\,\Cd\,
	\otimes \,e^{-[\ETA{C}^2 + \ETA{L}^2]/2}
	\sum_{n,m=0}^\infty \frac{(i\ETA{C})^{2n}}{n! n!}
	\frac{(i\ETA{L})^{2m}}{m! m!} \Ad{y}^n \A{y}^n \Ad{x}^m \A{x}^m
   + {\rm H.c.}\,,
 \label{mygun6}
\end{eqnarray}
\noindent
\hspace*{\fill}
\rule[0.4pt]{0.4pt}{\baselineskip}\hspace{-0.4pt}\rule[\baselineskip]{0.5\textwidth}{0.4pt}
\begin{multicols}{2}
\noindent
where the only time dependence stems from the envelope of the external
laser pulse. In writing \ceq{\ref{mygun6}} we have further assumed
that the corresponding pulse turn-on and off is slow when compared
with the trap frequencies $\Nu{x}$ and $\Nu{y}.$ This implies that the
duration of the laser pulse needs to be long when compared with
$\Nu{x,y}^{-1},$ to avoid the excitation of vibrational sidebands
through frequency components within the bandwidth of the pulse. In
\ceq{\ref{mygun6}} we have introduced the Lamb-Dicke parameters
$\ETA{L} = \klaser\dx$ and $\ETA{C} = \kcav\dy$ for the transitions
that are coupled through the laser and cavity electric fields,
respectively. In the Lamb-Dicke limit \cite{Lamb-Dicke,blockley92},
$\ETA{L,C} \ll 1,$ and we keep only the lowest order terms in the
Lamb-Dicke parameters to obtain
\begin{eqnarray}
 \HI(t) &=& -\hbar \delta_\ground\,\rground \lground\,\Cd\C 
	-\hbar \delta_\excited(t)\,\rexcited\lexcited
	\nonumber \\
    &&- \hbar g(t)\,\rground\lexcited\,\Cd
	+ {\rm H.c.}\,,
 	\label{mygun7}
\end{eqnarray}
so that the motional degrees of freedom decouple and the system
depicted in \fig{fig5} reduces to the scheme of Gheri {\em et al.\/}
where only the electronic and cavity degrees of freedom participate in
the dynamics. 

In the bad-cavity regime the cavity decay rate $\kappa$ sets the
fastest time scale in the system and dominates the dynamics of the
$\rground \Leftrightarrow \rexcited$ transition inhibiting any Rabi
oscillations between the states $\rground$ and $\rexcited.$ In this
limit, given that the initial state of the ion-cavity system is
$\rstate = \rexcited \rvib{0}_C$, and the coherent system dynamics are
described by the Hamiltonian in \ceq{\ref{mygun7}}, Gheri {\em et
al.\/} derive the solution
\begin{eqnarray*}
 \rstatet &=& \exp{\left[- \int_0^t \left( \frac{|g(\tp)|^2}{\kappa} 
	- i \delta_\excited(\tp) \right) d\tp \right]}
	\nonumber \\
	& & \times \left( \rexcited\rvib{0}_C
 	+ i \frac{g(t)}{\kappa} \rground\rvib{1}_C \right)\,,
\end{eqnarray*}
for the ion-cavity state {\em before\/} the cavity excitation has been
lost to the external field and where $\rvib{n}_C$ denote the usual
number states for the electric field inside the cavity. With this
result the probability of detecting a single photon during the time
interval $[0,t]$ can be calculated as $P(t) = 2\kappa \int_0^t
\lstatetp \Cd \C \rstatetp d\tp,$ if an ideal photodetector is used
\cite{law97}. From this we obtain the expression in
\ceq{\ref{mygun1}}, where the time dependence is essentially
controlled through the triggering laser pulse.

We have already mentioned that for a near perfect projective
measurement of the ion's electronic degrees of freedom we require
\begin{equation}
 {\rm Arg} \equiv 2 \int_0^\Tl \frac{|g(t)|^2}{\kappa} dt \gg 1\,,
 \label{mygun9}
\end{equation}
for the argument of the exponential term in \ceq{\ref{mygun1}}. This
can be achieved through a sufficiently large pulse area or pulse
duration which we have denoted by $\Tl$ here. To further elaborate on
this condition we consider again the specific example of $\rm Be^+$
ions as indicated in \fig{fig2}(b). For the value of the cavity decay
rate and the ion-cavity coupling strength we assume $\kappa/2\pi = 750
\rm kHz$ and $|\gcav|/2\pi = 5 \rm MHz,$ which are experimentally
realistic parameters \cite{parameters}. To remain specific we model
the explicit time dependence of the laser-ion coupling through the
pulse shape given in \ceq{\ref{pulseshape}} so that $\glaser(t) = f(t)
\glaser$,
and assume $|\glaser|/2\pi = 5 \rm MHz$, for the peak value of the
corresponding coupling constant. In view of the experiment reported in
Ref.\cite{meekhof96} this is a moderate assumption. For the value of
the detuning of the laser and cavity electric field from the excited
state $\rraman$, we assume $|\Delta|/2\pi = 250 {\rm MHz} \gg
|\glaser|, |\gcav|,$ so that the condition for the adiabatic
elimination of the excited state is satisfied. The parameters for
$|\glaser|, |\gcav|$ and $|\Delta|$ lead to $|g|/2\pi = 100 \rm kHz,$
for the peak value of the Raman coupling constant and the assumption
of the bad-cavity limit where $|g| \ll \kappa$ is
justified. Furthermore, this value for $|g|$ and the value of the
Stark shifts induced by the laser pulse and the cavity electric field,
$|\delta_{\ground,\excited}|/2\pi = 100 \rm kHz$, are well below the
typical trap frequencies $\Nu{x,y}/2\pi \approx 10 \rm MHz,$ in
experiments with $\rm Be^+$ \cite{meekhof96}, so that the assumption
of the low-excitation regime is valid. With these parameters and the
expression for the laser pulse shape given in \ceq{\ref{pulseshape}}
the argument of the exponential term in \ceq{\ref{mygun1}} can be
calculated explicitely and the condition in \ceq{\ref{mygun9}} imposes
the limit
\begin{equation}
 \Tl \gg \frac{4 \kappa}{3 |g|^2} \approx 16 \mu s\,,
 \label{reduce3}
\end{equation}
on the duration of the external laser pulse. This condition also
satisfies the requirement $\Tl \gg \Nu{x,y}^{-1} \approx 16 ns,$ so
that the bandwith of the pulse is small when compared with the
separation of the vibrational sidebands. For $\Tl = 100 \mu s,$ the
probability in \ceq{\ref{mygun1}}, $P(\Tl) = 99.8\%,$ and the
detection of the photon transmitted from the cavity represents a near
perfect projective measurement of the ion in its electronic state
$\rstatee = \rexcited.$ Thus, through the introduction of a low-Q
optical cavity we are able, in a particular limit, to read out the
error syndrome which is stored in the electronic ground states without
disturbance of the motional state. The duration of the projective
measurement can be shortened for larger values of the Raman coupling
constant $|g|$, which may be achieved through an increase of the
electric field of the triggering laser pulse, if at the same time the
trap frequencies can be increased to remain in the low-excitation
regime.

\section{The Unitary Inversion}
\label{repair}

In Sections \ref{dissipation} and \ref{detection} we have described
how to encode the qubit in stable motional basis states and how to
detect the quantum jumps associated with the master equation given in
\ceq{\ref{140}}. We now turn to the unitary inversion of the detected
quantum jumps, the processes which constitute step (ii) of our
stabilisation scheme as depicted in \fig{fig1}. In the case where no
quantum jump has been detected through the interrogation process
described in Section \ref{detection} the ion is left in the motional
state given in \ceq{\ref{nojstate}} which is identical to the initial
motional state in \ceq{\ref{encoding}}, so that no further
manipulation is required to restore the quantum information to the
original basis states. On the other hand, the detection of a quantum
jump associated with the decay channel $\jump{x,y}$ results in the
state
\begin{equation}
 \rstatev_{x,y} = c_- \basism_{x,y} + c_+ \basisp_{x,y}\,,
 \label{Jumpstate}
\end{equation}
corresponding to the channel operators $\jump{x}$ and $\jump{y},$ and
the states $\basispm_{x,y}$ have been given in Eqs.\,(\ref{Xbasis})
and (\ref{Ybasis}). The unitary restoration of the quantum information
then requires the transformation
\begin{equation}
 \hat{U}_{\it rest}^{x,y} \basispm_{x,y} = \sbasispm\,.
 \label{unitaryrestore}
\end{equation}
The generation of this transformation is not simple. The best solution
would be to directly generate the unitary operator in
\ceq{\ref{unitaryrestore}} by ``dialing up" a specific finite unitary
operator which acts within the two-dimensional motional Hilbert space.
Although there has been some discussion concerning the generation of
any given finite unitary operator \cite{Reck}, this has yet to be
implemented for the states of motion of a trapped ion. We were able to
find a discrete, finite two-mode unitary transformation which will
effect the restoration in \ceq{\ref{unitaryrestore}} \cite{jason}, but
we know of no systematic physical mechanism which could implement this
transformation. In the absence of such a mechanism we are forced to
look for a particular sequence of processes that will effect the
unitary inversion of the quantum jumps associated with the decay
channels $\jump{x}$ and $\jump{y}$ as in
\ceq{\ref{unitaryrestore}}. We will attempt to avoid the use of
resonant interactions as they are prey to severe timing
constraints. Instead, we will effect the restoration using two
adiabatic transfer processes and a single, intermediate stimulated
process.

We will only describe the procedure which accomplishes the
transformation in \ceq{\ref{unitaryrestore}} for the case where a
quantum jump via the $\jump{x} = \sqrt{\gamma} \A{x}$ channel has been
detected. The case for the $\jump{y} = \sqrt{\gamma} \A{y}$ channel is
almost identical. After the detection of a quantum jump associated
with the decay channel $\jump{x}$, the ion is left in the product state
$\rstate_x = \rstatev_x \otimes \rground,$ where the motional state is
\begin{equation}
  \rstatev_x = \frac{1}{\sqrt{2}} (c_+ + c_-)\rtwomode{3}{0}
  + (c_+ - c_-)\rtwomode{1}{2}\,,
 \label{start}
\end{equation}
and where the coefficients $c_\pm$ are {\it unknown}, as seen from
Eqs.\,(\ref{xjstate}) and (\ref{Xbasis}). The first step in the
restoration of the state $\rstatev_x$ to the original state $\rstatev$
will be to coherently add one quantum to the vibrational excitation in
the $x$ direction. To accomplish this we make use of adiabatic passage
techniques \cite{schneider98b}, which have previously been studied in
the context of coherent population transfer \cite{vitanov97}. We
coherently transfer population between the electronic states
$\rground$ and $\rexcited$ through a pair of overlapping, time-delayed
laser pulses which are known as the pump and the Stokes pulse
\cite{vitanov97}. More specifically, we assume that the laser which
generates the pump pulse is resonant with the $\rground
\Leftrightarrow \rraman$ transition, and the laser that generates the
Stokes pulse is tuned to the first red sideband of the $\rexcited
\Leftrightarrow \rraman$ transition, and aligned with the $x$ axis. As
shown in Appendix \ref{japp}, this excitation of the $\Lambda$ system,
depicted in \fig{fig2}, is described by the Hamiltonian
\begin{equation}
 \H_{\it add} = -\hbar \ggr(t) \rground\lraman
  -\hbar \ger(t)\Ad{x} \otimes \rexcited\lraman 
  + {\rm H.c.}\,,
	\label{Hadd}
\end{equation}
where the time dependence of the coupling constants $\ggr(t)$ and
$\ger(t)$ stems from the dimensionless shapes $f_\ground(t)$ and
$f_\excited(t)$ of the pump and the Stokes pulse as seen from
\ceq{\ref{adiabcoupl}}. The above Hamiltonian possesses the particular
instantaneous eigenstate
\begin{eqnarray}
 && \rpsi{n_x n_y}_{\it dark}
	\nonumber \\
 &&= \frac{ \alpha(t) \rtwomode{n_x}{n_y} \otimes \rground 
 - \beta(t) \rtwomode{n_x+1}{n_y} 
	\otimes \rexcited}{\sqrt{\alpha(t)^2 + \beta(t)^2}}\,,
 \label{adddark}
\end{eqnarray}
whose corresponding eigenvalue vanishes. This eigenstate does
not contain any contribution from the excited state $\rraman$, 
and is generally known as a ``dark
state" \cite{alzetta76}. The explicit time dependence of this state
is given through the quantities $\alpha(t) = \ger^*(t) [(n_x+1)! /
n_x!]^{1/2}$ and $\beta(t) = \ggr^*(t).$ The addition of a single
quantum in the vibrational excitation along the $x$ direction is
accomplished through adiabatically following the dark state in a
counter-intuitive pulse sequence, where the Stokes pulse overlaps but
precedes the pump pulse \cite{vitanov97}. In this situation the dark
state $\rpsi{n_x n_y}_{\it dark} = \rtwomode{n_x}{n_y} \otimes
\rground,$ before the pump pulse is turned on, and $\rpsi{n_x
n_y}_{\it dark} = \rtwomode{n_x+1}{n_y} \otimes \rexcited,$ after the
Stokes pulse is turned off, independent of the vibrational
excitation numbers $n_x$ and $n_y$. If the pulse sequence is performed
adiabatically the system follows the dark state given above and, after
the adiabatic passage, the ion is left in the state
\begin{eqnarray}
  \rstate_x^\prime &=& 
	\frac{1}{\sqrt{2}} \left\{ (c_+ + c_-)\rtwomode{4}{0}
	\right. \nonumber \\
 && \left.
      + (c_+ - c_-)\rtwomode{2}{2}\right\} \otimes \rexcited\,.
\label{leftwitha}
\end{eqnarray}
This is accomplished without severe constraints on the pulse durations
and amplitudes, as long as the process is performed adiabatically. The
adiabaticity condition is satisfied if
\begin{equation}
 \int_{0}^T \left| g_{\ground,\excited}(t) \right| dt \gg 1 \,,
 \label{adiabaticity}
\end{equation}
where $T$ is the duration of the adiabatic passage \cite{vitanov97}.
In addition we note that in the adiabatic limit, the excited state
$\rraman$ does not get populated during the entire process, so that
spontaneous emission plays no role.

In the second step of our unitary restoration of the quantum
information to the original basis states we coherently split the
electronic population associated with the motional state
$\rtwomode{4}{0},$ in the superposition of \ceq{\ref{leftwitha}},
without affecting the state $\rtwomode{2}{2}.$ This is accomplished
through a unitary time evolution governed by the Hamiltonian
\begin{equation}
  \H_{\it split}= \hbar g(t) \left\{ \Ad{x}\A{x} -\Ad{y}\A{y} \right\}
  \otimes \rground\lexcited +{\rm H.c.} \,, 
	\label{flip}
\end{equation}
whose generation through two pairs of lasers that drive a stimulated
Raman transition between the electronic states $\rground$ and
$\rexcited$, follows closely the manner in which the Hamiltonian given
in \ceq{\ref{hamI}} was constructed. The first pair of Raman lasers is
arranged so that the wavevector difference $\underline{\delta
k}^{(1)}$ is aligned with the $x$ axis, whereas, for the second pair
of Raman lasers, we assume that the wavevector difference
$\underline{\delta k}^{(2)}$ is now aligned with the $y$
axis. Following \ceq{\ref{HRaman}}, this gives rise to the
Hamiltonians
\begin{eqnarray}
 \H^{(1,2)} &=& - \hbar \gge^{(1,2)}(t)
	 \exp{\left[-\eta^2/2 \right]}
	 \,\sm \nonumber \\
    && \otimes \left\{ {\bf 1} - \eta^2 \Ad{x,y}\A{x,y} \right\}
	+ {\rm H.c.}\,,
 \label{hamxy}
\end{eqnarray}
for the first and the second pair of Raman lasers, respectively. Here,
we have further assumed the Lamb-Dicke limit
\cite{Lamb-Dicke,blockley92}, and $\ETA{x}^{(1)} = \ETA{y}^{(2)} =
\eta,$ for the Lamb-Dicke parameters, which can be accomplished through
an appropriate choice of $|\underline{\delta k}|^{(1)}$ and
$|\underline{\delta k}|^{(2)}$ as seen from \ceq{\ref{ldp}}. Given
that the laser phases are arranged such that $g^{(1)}(t) = -
g^{(2)}(t),$ the dynamics generated by the combination of the two
pairs of Raman lasers is described by the Hamiltonian in
\ceq{\ref{flip}}, and the coupling constant is given by
\ceq{\ref{couplingconst}}. The resulting time evolution is identical
to the one in \ceq{\ref{unitaryI}}, but where now $\hat{\chi} = |g|
(\Ad{x}\A{x} - \Ad{y}\A{y})$. With the particular choice $A = \pi / 16
|g|,$ for the generalized pulse area in \ceq{\ref{395}}, the
Raman-induced dynamics leave the ion in the state
\begin{eqnarray}
  \rstate_x^{\prime\prime}&=& 
	\frac{1}{2} \big\{ (c_+ + c_-)\rtwomode{4}{0} 
	\otimes (\rground + \rexcited)  \nonumber \\
 &&  + (c_+ - c_-)\sqrt{2}\rtwomode{2}{2}\otimes \rexcited \big\} \,,
\label{leftwithb}
\end{eqnarray}
after the second step in our unitary restoration.

In the third and final step, we complete the transformation in
\ceq{\ref{unitaryrestore}} through a second adiabatic transfer process
which is similar to the one that we have employed in the first step.
Starting from the state given in \ceq{\ref{leftwithb}}, we recombine
the electronic population in the state $\rexcited$ by adiabatically
transferring the population from the electronic state $\rground$ to
the state $\rexcited$, such that the motional excitation is
simultaneously transferred from the $x$ into the $y$ direction. This
is done through an adiabatic passage in counter-intuitive pulse
sequence as described above but where we now assume that the laser
which generates the pump pulse is aligned with the $x$ axis and tuned
to the fourth red sideband of the $\rground \Leftrightarrow \rraman$
transition. The laser that generates the Stokes pulse is tuned to the
fourth red sideband of the $\rexcited \Leftrightarrow \rraman$
transition and aligned with the $y$ axis. As discussed in Appendix
\ref{japp}, this leads to the Hamiltonian
\begin{eqnarray}
 \H_{\it comb} &=& -\hbar \ggr(t)\Ad{x}^{\!4}
	\otimes \rground\lraman
	\nonumber \\
  && -\hbar \ger(t)\Ad{y}^{\!4}\otimes \rexcited\lraman 
  + {\rm H.c.}\,,
	\label{Hcombine}
\end{eqnarray}
which has the important property that $\hat{U}_{\it comb}\equiv 
e^{-i\hat{H}_{\it comb}t/\hbar} \equiv {\bf 1}$ on the states
$\rtwomode{4}{0}\otimes\rexcited$ and $\rtwomode{2}{2} \otimes
\rexcited.$ The only component in the superposition state of
\ceq{\ref{leftwithb}} which is affected by the adiabatic transfer
is the state $\rtwomode{4}{0} \otimes \rground.$ This state coincides
with a component of the dark state associated with the above Hamiltonian
\begin{eqnarray}
 &&\rpsi{n_x n_y}_{\it dark} 
 \nonumber \\
 &&= \frac{ \alpha(t) \rtwomode{n_x+4}{n_y} \otimes \rground 
 - \beta(t) \rtwomode{n_x}{n_y+4} 
	\otimes \rexcited}{\sqrt{\alpha(t)^2 + \beta(t)^2}}\,,
	\nonumber \\
 && \label{combinedark}
\end{eqnarray}
for $n_x = n_y =0,$ before the pump pulse is turned on. As before, the
time dependence of the dark state is given through the quantities
$\alpha(t) = \ger^*(t) [(n_y+4)!/n_y!]^{1/2}$ and $\beta(t) =
\ggr^*(t) [(n_x+4)!/n_x!]^{1/2}$, so that after the Stokes pulse is
turned off, the state $\rtwomode{4}{0} \otimes \rground$ has been
adiabatically transferred to the state $\rtwomode{0}{4} \otimes
\rexcited.$ The adiabatic process generated by the pump and the Stokes
pulse then leaves the ion in the product state $\rstate = \rstatev
\otimes \rexcited,$ where the motional state
\begin{eqnarray}
  \rstatev &=& \frac{1}{2} \big\{ (c_+ + c_-) (\rtwomode{4}{0} 
	+ \rtwomode{0}{4} )  \nonumber \\
 &&  + (c_+ - c_-)\sqrt{2}\rtwomode{2}{2}\big\}\,,
\label{leftwithc}
\end{eqnarray}
and where we have chosen the two lasers beams to be $\pi$ out of phase so
as to cancel the accumulated phase acquired by $\rtwomode{0}{4}
\otimes \rexcited.$ The final state in \ceq{\ref{leftwithc}} 
is identical to the initial motional state in \ceq{\ref{encoding}},
and thus the stable motional basis states given in \ceq{\ref{mystates}}
have been completely restored.

To summarize the various steps in the above proposed restoration of
the qubit after the detection of a quantum jump associated with the
$\jump{x}$ channel: we first adiabatically increased the excitation
number of the ion's motion along the $x$ direction using two
time-delayed laser pulses. Secondly we coherently split the electronic
population associated with the motional state $\rtwomode{4}{0}$ into
the superposition $\rground + \rexcited$ using a unitary
transformation which we generated through a stimulated Raman process
induced by two pairs of laser beams. The third step of the restoration
sequence then served to establish the superposition $\rtwomode{4}{0} +
\rtwomode{0}{4}$ in the motional basis states 
and at the same time disentangled the ion's motional and electronic
degrees of freedom. This final step was achieved through a second
adiabatic transfer process generated by two time-delayed laser
pulses. 

To complete our analysis of the time scales associated with our
stabilisation scheme, we examine here the adiabatic transfer
processes involved in the unitary inversion of the detected quantum
jumps. From our discussion at the end of Section
\ref{entanglement} we can estimate the duration of the laser pulse
that generates the intermediate, resonant step which led us from
\ceq{\ref{leftwitha}} to \ceq{\ref{leftwithb}}, to be $\Tl \approx 8
\mu s.$ The minimal duration of the adiabatic transfer steps is
limited by the adiabaticity condition in
\ceq{\ref{adiabaticity}}. This is essentially controlled through the
magnitude of the coupling constants $|g_\ground|$ and
$|g_\excited|$. We consider here the second adiabatic transfer process
since this involves the excitation of the ion's fourth vibrational
sidebands which is highly suppressed through the smallness of the
Lamb-Dicke parameters $\ETA{x,y}$. More specifically
$|g_{\ground,\excited}| = \ETA{x,y}^4
|\tilde{g}_{\ground,\excited}|/4!$, as seen from
\ceq{\ref{adiabcoupl}} where $|\tilde{g}_\ground|$
($|\tilde{g}_\excited|$) gives the resonant coupling strength for the
dipole transition between the states $\rground$ ($\rexcited$) and
$\rraman$. The magnitude of the resonant coupling strengths is in turn
limited by the off-resonant excitation of other sidebands and in
particular the carrier transition which we have neglected in making
the vibrational rotating wave approximation. The mathematical
condition for this limit is $|\tilde{g}_{\ground,\excited}| / 4
\Nu{x,y} \ll 1$ \cite{Ciracadv}, and for $\Nu{x,y}/2\pi = 10 \rm MHz,$
imposes $|\tilde{g}_{\ground,\excited}| / 2\pi \ll 40 \rm MHz,$ on the
magnitude of the resonant coupling strengths. We assume
$|\tilde{g}_{\ground,\excited}| / 2\pi = 15 \rm MHz,$ which for
$\ETA{x,y} = 0.2,$ leads to $|g_{\ground,\excited}| = 1 \rm kHz,$ for
the coupling strength corresponding to the excitation of the fourth
vibrational sidebands. With this result the adiabaticity condition in
\ceq{\ref{adiabaticity}} imposes the limit 
\begin{equation}
 \Tl \gg 320 \mu s\,,
 \label{limit3}
\end{equation}
on the duration of the pulses that generate the adiabatic transfer and
where we have assumed the pulse shape given in
\ceq{\ref{pulseshape}}. The duration of this process may be shortened
for larger values of the resonant coupling strengths
$|\tilde{g}_{\ground,\excited}|$, which would however require a
simultaneous increase in the trap frequencies so as to avoid the
off-resonant excitation of other than the chosen sidebands.

Following our assessment for the duration of the various
processes involved in our stabilisation scheme (Eqs.\,(\ref{limit1}),
(\ref{reduce3}) and (\ref{limit3})), we can estimate the total
duration of steps (i) and (ii) indicated in \fig{fig1} to be of the
order of $1 ms,$ with currently available technology. With this result
we can give a figure of merit for the performance of our scheme. We assume
$\tau = 10 ms,$ for the period with which we interrogate the
dissipative system evolution and $\gamma = 0.1 s^{-1}$ for the
decoherence rate. With these values the probability for two quantum
jumps to occur during the time interval $\tau$ equals
the probability for a single quantum jump to occur during the
interrogation process and is of the order of $10^{-2}$. Our
stabilisation scheme then suppresses the rate of decoherence by two
orders of magnitude. We note that this is not a fundamental limit and
can be improved through reducing the duration of the operations that
constitute step (i) and (ii) in \fig{fig1}. As discussed above this
would primarily require an increase in the trap frequencies
$\Nu{x,y},$ beyond current laboratory values. 


\section{Conclusion}
\label{conclusion}

To conclude, we have considered the active stabilisation of a qubit
which is stored in the bosonic motional degrees of freedom of an
ultra-cold ion and which is suffering a particular type of motional
decoherence. Through a quantum-trajectory analysis of the dissipative
system evolution associated with a finite-temperature master equation
we were able to uncover particular basis states with which to encode
the qubit and which obey the necessary criteria for the effect of a
single quantum jump and the absence of quantum jumps associated with
this form of decoherence to be unitarily reversible.  We found that
there exists a duality between the different unravelings of the master
equation and the form of the stable basis states associated with each
unraveling. Based on this duality we chose a form for the basis states
such that the signature of the quantum jumps can be extracted through
two consecutive binary interrogations. These interrogations determine
if a single jump has taken place and if so, which type of jump. The
final stage of the interrogation involved the reading out of the
information held in the electronic states. We proposed a method which,
through coupling the ion to a low-Q optical resonator, can transfer
this electronic information to the environment with very little
disturbance to the motion. Finally, the unitary inversion of the
detected quantum jumps required two adiabatic processes and one
stimulated process.  The entanglement (Section \ref{entanglement}),
and unitary inversion (Section \ref{repair}), steps utilized, at most,
two pairs of Raman lasers. This should be experimentally feasible.
The repair operation described here is completely unitary, and does
not make recourse to any non-unitary probabilistic operations. Thus it
should be ideally suited to operate on single systems such as the ions
in a quantum computer. In addition to being stable against the effects
of a zero-temperature bath, we have shown that the particular encoding
found is also stable against the effects of a thermal bath. This
result might prove very significant as thermal noise plays a large
role in the systematic sources of decoherence in trapped ions
\cite{heating}. However, we have restricted our discussion regarding
the quantum jump detection and inversion processes to the case of a
zero-temperature bath. The case of a thermal bath will be better
formulated when the separate issue of the generation of any given
motional unitary operator has been solved.  The entire process of
restoration was made possible by the {\it a priori} knowledge of the
type of dissipation present. In a typical experiment, there will be
many varied sources of noise and for the above stabilisation to be
useful, information concerning the various sources decoherence must be
experimentally obtained.  Finally, it would be very interesting to
learn whether the above formal analysis concerning the existence of
stable bases extends to more complicated types of dissipation. The
practical implementation of such extensions will, however, pose a
formidable task.

\section*{Acknowledgments}

This work was supported in part by the UK Engineering and Physical
Sciences Research Council and the European Community. J.T.\ thanks the
Royal Society and the Royal Irish and Austrian Academies for
support. J.S.\ is supported by the German Academic Exchange Service
(DAAD-Doktorandenstipendium aus Mitteln des dritten
Hochschulsonderprogramms).  The authors would like to thank M.B.~Plenio,
P.L.~Knight, V.~Bu\v{z}ek, P.~Zoller, I.~Cirac and D.M.~Segal for useful
discussions.

\begin{appendix}
\section{Pulsed Stimulated Raman transitions}
\label{ramanapp}

The transition between the two ground state hyperfine levels
$\rground$ and $\rexcited$ can be driven by two lasers connecting
these two states to a common excited state $\rraman,$ in a stimulated
Raman scheme \cite{monroe95,steinbach97}, as indicated in
\fig{fig2}. We briefly review this here for the case of a pulsed
transition and derive some of the notation necessary for our
discussion in Section \ref{entanglement}. The stimulated Raman
transition is driven by the electric field
\begin{eqnarray}
 E(t) &=& E_{\ground}(t) \,
			e^{-i[\k{\ground}.\R - \omegagr t]}
		     + {\rm c.c.}
 \nonumber \\
		 &&  + \  E_{\excited}(t) \,
			e^{-i[\k{\excited}.\R - \omegaer t]}
		     + {\rm c.c.\,,}
 \label{400}
\end{eqnarray} 
where $E_{\ground}(t) = E_{\ground} f(t),$ and $E_{\excited}(t) =
E_{\excited} f(t)$ are the electric field amplitudes of the two lasers
and $f(t)$ is a dimensionless pulse shape, which we assume to be the
same for both lasers. In \ceq{\ref{400}} we have denoted the frequency
of the laser driving the $\ri \Leftrightarrow \rraman$ transition by
$\omegair,$ and the corresponding wavevector is given by
$\underline{k}_{\i},$ where $\i = \ground,\excited$. To generate the
stimulated Raman transition between the two states $\rground$ and
$\rexcited$ without populating the upper level $\rraman,$ we assume
both lasers to be far detuned from the excited state $\rraman$ and
denote the corresponding detuning by $\Deltair = (\omraman -
\omi)-\omegair.$ We are considering a {\em resonant\/} two-photon
transition between the states $\rground$ and $\rexcited$ so that
\begin{equation}
 \Deltagr = \Deltaer = \Delta\,, 
 \label{ramandet}
\end{equation} 
for the detuning of the two lasers from the excited state $\rraman$.
In the dipole approximation, and after performing the optical rotating
wave approximation, we can adiabatically eliminate the excited state
$\rraman$ under the conditions $|\Delta| \gg |\ggr(t)|,|\ger(t)|,$ as
shown in \cite{steinbach97}. Here we have defined the dipole coupling
constants $\gir(t) = f(t)\,\gir = f(t)\,\li \hat\wp \rraman
E_{\i}/\hbar,$ and $\hat\wp$ is the dipole operator. After the
adiabatic elimination we obtain the Hamiltonian
\begin{eqnarray}
 \H &=& - \hbar \deltag(t)\,\rground\lground
	- \hbar \deltae(t)\,\rexcited\lexcited \nonumber \\[2ex]
    && - \hbar \gge(t)
	 \exp{\left[-\left(\ETA{x}^2+\ETA{y}^2\right)/2\right]}
	 \,\sm \nonumber \\
    && \otimes \prod_{j=x,y}\,\sum_{n_j,m_j=0}^\infty 
	 \frac{(-i\ETA{j})^{m_j+n_j}}{m_j! n_j!} \Ad{j}^{m_j} \A{j}^{n_j} 
    \nonumber \\[2ex]
    && \ \ \ \ \ \times \exp{\left[-i\,\nu_j\left(n_j-m_j\right)t \right]} 
	+ {\rm H.c.}\,,
 \label{410}
\end{eqnarray}
in the interaction picture of $\HO,$ which was given in
\ceq{\ref{20}}. Here we have introduced the Raman coupling constant
\begin{equation}
 \gge(t) = f(t)^2 \gge\,,
 \label{411}
\end{equation}
where $\gge = \ggr \ger^* /2 \Delta,$ and defined the Lamb-Dicke
parameters $\ETA{x} = \Delta x_0\, \delta k_x$ and $\ETA{y} = \Delta
y_0\, \delta k_y,$  
through the $x$ and $y$ component of the wavevector difference
$\underline{\delta k} = \k{\ground} - \k{\excited},$ and the width of
the motional ground state along the $x$ and $y$ axes as given by
$\Delta x_0 = (\hbar/2\Nu{x}m)^{1/2}$ and $\Delta y_0 = (\hbar/
2\Nu{y}m)^{1/2}.$

The stimulated Raman transition has the disadvantage that it gives
rise to optical Stark shifts which are given by the terms
\begin{equation}
 \deltag(t) = \frac{2 |\ggr(t)|^2}{\Delta}\,, 
 \qquad
 \deltae(t) = \frac{2 |\ger(t)|^2}{\Delta}\,,  
 \label{415}
\end{equation}
and which are time dependent in the case of a pulsed transition
\cite{steinbach97}. If these shifts are equal, i.e.\ for $|\ggr(t)|^2
= |\ger(t)|^2,$ the relative energy shift between the states
$\rground$ and $\rexcited$ is constant and zero, and the optical Stark
shifts will merely contribute as an overall phase to the dynamics
generated by the Hamiltonian in \ceq{\ref{410}}. We will assume
$\deltag(t) = \deltae(t)$ and drop the contribution of these optical
Stark shifts, thereby not explicitly keeping track of overall phases
which are not important to us here. This situation can be
experimentally realized by varying the relative intensity of the two
Raman pulses
\cite{wineland98}. Alternatively, it has been shown that appropriately
chirped laser pulses can compensate for these time varying Stark
shifts \cite{manos98}.

In our discussion in Section \ref{entanglement} we are interested in
the situation where the stimulated Raman transition described by the
Hamiltonian in \ceq{\ref{410}} is only sensitive to the motion of the
ion along one of the principal axes. This can be realized by arranging
the two exciting laser beams so that their wavevector difference
$\underline{\delta k} = \k{\ground} - \k{\excited},$ is aligned with
that principal axis. The Hamiltonian given in \ceq{\ref{410}} then
simplifies to
\begin{eqnarray}
 \H &=& - \hbar \gge(t)
	 \exp{\left[-\ETA{j}^2/2\right]} \,\sm \nonumber \\
    && \otimes \sum_{n,m=0}^\infty 
	 \frac{(-i\ETA{j})^{m+n}}{m! n!} \Ad{j}^{m} \A{j}^{n} 
    \nonumber \\[2ex]
    && \ \ \times \exp{\left[-i\,\nu_j\left(n-m\right)t \right]} 
	+ {\rm H.c.}\,,
 \label{412}
\end{eqnarray}
where $j=x$ for the case of aligning $\underline{\delta k}$ with the
$x$ axis, and $j=y$ for the case of aligning $\underline{\delta k}$
with the $y$ axis.

A further simplification arises in the low excitation regime
\cite{Ciracadv}, and if the pulse turn-on and off is slow when
compared to the trap frequencies $\Nu{x}$ and $\Nu{y}.$ We can then
perform the vibrational rotating wave approximation \cite{blockley92},
and consider only the resonant terms in \ceq{\ref{412}}, to obtain
\begin{eqnarray}
 \H &=& - \hbar \gge(t)
	 \exp{\left[-\ETA{j}^2 / 2\right]}
	 \,\sm \nonumber \\
    && \otimes \sum_{n=0}^\infty 
	 \frac{(-i\ETA{j})^{2 n}}{n! n!} \Ad{j}^{n} \A{j}^{n} 
	+ {\rm H.c.}\,,
 \label{Hraman}
\end{eqnarray}
where $j=x,y$ as explained above. Without inducing transitions between
the ion's vibrational levels, this Hamiltonian is sensitive to
the ion's motional state through its dependence on the creation and
annihilation operators for the ion's vibrational excitation.

\section{Adiabatic Passage}
\label{japp}

To effect the unitary inversion of the decoherence processes
associated with the decay channels $\jump{x}$ and $\jump{y}$, we make
use of adiabatic passage techniques which have previously been studied
in the context of coherent population transfer \cite{vitanov97}, where
atomic population is coherently transferred via Raman transitions
induced by a pair of overlapping time-delayed laser pulses. These
techniques rely on the existence of a dark state of the corresponding
Hamiltonian. In Section \ref{repair} we employ adiabatic passage
processes that are based on the dark states associated with the two
types of Hamiltonians
\begin{eqnarray}
 \H_{\it dark} &=& -\hbar \ggr(t)\Ad{y}^{\!\kappa_{\ground}}
	\otimes \rground\lraman
	\nonumber \\
  && -\hbar \ger(t)\Ad{x}^{\!\kappa_{\excited}}\otimes \rexcited\lraman 
  + {\rm H.c.}\,,
	\label{HgeneralI}
\end{eqnarray}
and
\begin{eqnarray}
 \H_{\it dark} &=& -\hbar \ggr(t)\Ad{x}^{\!\kappa_{\ground}}
	\otimes \rground\lraman
	\nonumber \\
  && -\hbar \ger(t)\Ad{y}^{\!\kappa_{\excited}}\otimes \rexcited\lraman 
  + {\rm H.c.}\,,
	\label{HgeneralII}
\end{eqnarray}
which describe the excitation of the $\Lambda$ system depicted in
\fig{fig2} with two laser pulses that drive specific vibrational
sidebands of the dipole transitions $\rground \Leftrightarrow \rraman$
and $\rexcited \Leftrightarrow \rraman,$ respectively. Here, we
briefly describe the details of this excitation. Vogel and
de~Matos~Filho \cite{nonlinearJCM}, have shown that the excitation of
the dipole transition $\ri \Leftrightarrow \rraman,$ (where $\i =
a,b$) with an electric field
\begin{equation}
 E_\i(t) = f_i(t) E_{\i}\,e^{-i[\k{\i}.\R - \omegair t]} + {\rm c.c.}\,,
 \label{Edipole}
\end{equation}
which is tuned to the $\kappa$th red sideband of the transition is
described by a nonlinear $\kappa$-quantum Jaynes-Cummings model in the
form
\begin{eqnarray}
 \H &=& - \hbar \tilde{g}_\i(t)
	 \exp{\left[-\ETA{j}^2 / 2\right]}
	 \,\ri\lraman \nonumber \\
    && \otimes \sum_{n=0}^\infty 
	 \frac{(-i\ETA{j})^{2 n + \kappa}}{n! (n+\kappa)!} 
	\Ad{j}^{n+\kappa} \A{j}^{n} + {\rm H.c.}\,,
 \label{Hdipole}
\end{eqnarray}
where $j=x,$ for the case of aligning the wavevector $\k{\i}$ with the
$x$ axis, and $j=y,$ for the case of aligning $\k{\i}$ with the $y$
axis \cite{nonlinearJCM}. Here we have defined the dipole coupling
constants $\tilde{g}_\i(t) = f_\i(t)\,\tilde{g}_\i = f_\i(t)\,\li
\hat\wp\rraman E_{\i}/\hbar,$ where $\hat\wp$ is the dipole operator
and the dimensionless quantity $f_\i(t)$ gives the time dependence of
the electric field amplitude as in \ceq{\ref{Edipole}}. The Lamb-Dicke
parameters are given by $\ETA{x} = \Delta x_0 |\k{\i}|$ and $\ETA{y} =
\Delta y_0 |\k{\i}|$ for the case of aligning the wavevector with the
$x$ and $y$ axis, respectively. In the Lamb-Dicke limit $\ETA{x,y} \ll
1$ \cite{Lamb-Dicke,blockley92}, and we retain only the leading order
term in the Lamb-Dicke parameter to obtain
\begin{equation}
 \H = - \hbar \tilde{g}_\i(t) e^{-\ETA{j}^2 / 2} 
	\frac{(-i\ETA{j})^\kappa}{\kappa!}
	\,\Ad{j}^\kappa \otimes \ri\lraman + {\rm H.c.}\,.
\label{Hldl}
\end{equation}
With this result, the pulsed excitation of the $\Lambda$ system with
two laser beams is described by the Hamiltonian given in
\ceq{\ref{HgeneralI}} if the laser that drives the $\rground
\Leftrightarrow \rraman$ transition is aligned with the $y$ axis and
tuned to the $\kappa_\ground$th red sideband, and the laser that
drives the $\rexcited \Leftrightarrow \rraman$ transition is aligned
with the $x$ axis and tuned to the $\kappa_\excited$th red
sideband. The Hamiltonian in \ceq{\ref{HgeneralII}} is obtained for
the same sideband detunings but where the laser that drives the $\rground
\Leftrightarrow \rraman$ transition is aligned with the $x$ axis, and
the laser that drives the $\rexcited \Leftrightarrow \rraman$
transition is aligned with the $y$ axis. The coupling constants in
Eqs.\,(\ref{HgeneralI}) and (\ref{HgeneralII}) are given by
\begin{equation}
 \gir(t) = f_\i(t) \gir = \tilde{g}_\i(t) \exp{\left[-\ETA{i}^2 / 2\right]} 
	\frac{(-i\ETA{i})^{\kappa_\i}}{\kappa_\i!}\,,
 \label{adiabcoupl}
\end{equation}
where $\i=\ground,\excited,$ and the Lamb-Dicke parameters $\ETA{a,b}
= \ETA{y,x}$ in \ceq{\ref{HgeneralI}}, and $\ETA{a,b} = \ETA{x,y}$ in
\ceq{\ref{HgeneralII}}. The time delayed pulses are characterized by
$f_\ground(t)$ and $f_\excited(t)$ which give the dimensionless pulse
shape.

\end{appendix}

\end{multicols}


\begin{thebibliography}{99}
\bibitem{harmonic} In treating the motion of the ion as harmonic in a
 static potential we are neglecting the effects of micromotion, whose
 main contribution is the alteration of the transition rates between
 quantum levels \cite{bardroff96}, which can be accounted for by
 experimental calibration \cite{wineland98}.
\end{thebibliography}
\end{document}